\documentclass[%
 reprint,
 amsmath,amssymb,
 aps,
nofootinbib
]{revtex4-2}

\usepackage{graphicx}
\graphicspath{{img/}}
\usepackage{dcolumn}
\usepackage{bm}
\usepackage{mathptmx}
\usepackage{hyperref}
\usepackage{xcolor}
\usepackage{bbold}
\definecolor{darkblue}{rgb}{0, 0, 0.6}
\hypersetup{
    colorlinks=true,
    linkcolor=darkblue,
    filecolor=darkblue,      
    urlcolor=darkblue,
    citecolor = darkblue,
    pdftitle={Extreme mass ratio inspirals into black holes surrounded by matter},
    pdfpagemode=FullScreen,
    }

\usepackage{subcaption}

\renewcommand{\d}{\mathrm{d}}
\renewcommand{\th}{\vartheta}
\newcommand{\ph}{\varphi}

\begin{document}

\title{Extreme mass ratio inspirals into black holes surrounded by matter} 

\author{Luk\'{a}\v{s} Polcar$^{1,2}$}
\author{Georgios Lukes-Gerakopoulos$^1$}
\email{gglukes@gmail.com}
\author{Vojt\v{e}ch Witzany$^3$ }
\affiliation{${}^1$ Astronomical Institute of the Czech Academy of Sciences, Bo\v{c}n\'{i} II 1401/1a, CZ-141 00 Prague, Czech Republic}   
\affiliation{${}^2$Institute of Theoretical Physics, Faculty of Mathematics and Physics, Charles University in Prague, 18000 Prague, Czech Republic}
\affiliation{${}^3$ School of Mathematics and Statistics, University College Dublin, Belfield, Dublin 4, D04 V1W8, Ireland}

\begin{abstract}
    Inspirals of stellar-mass compact objects into massive black holes, known as extreme mass ratio inspirals (EMRIs), are one of the key targets for upcoming space-based gravitational-wave detectors. In this paper we take the first steps needed to systematically incorporate the effect of external gravitating matter on EMRIs. We model the inspiral as taking place in the field of a Schwarzschild black hole perturbed by the gravitational field of a far axisymmetric distribution of mass enclosing the system. We take into account the redshift, frame-dragging, and quadrupolar tide caused by the enclosing matter, thus incorporating all effects to inverse third order of the characteristic distance of the enclosing mass. Then, we use canonical perturbation theory to obtain the action-angle coordinates and Hamiltonian for mildly eccentric precessing test-particle orbits in this background. Finally, we use this to efficiently compute mildly eccentric inspirals in this field and document their properties.
    This work shows the advantages of canonical perturbation theory for the modeling EMRIs, especially in the cases when the background deviates from the standard black hole fields.
\end{abstract}

\maketitle

\section{Introduction}

Extreme mass ratio inspirals (EMRIs) are one of the most complex sources of gravitational waves that we expect to be observed by the Laser Interferometer Space Antenna (LISA) \cite{EMRIsLISA}. These binary sources are composed of a primary supermassive black hole and a secondary much lighter compact object, such as a black hole or a neutron star. These systems are called extreme mass ratio (EMR), because the mass ratio between the secondary and the primary is below $10^{-4}$. Such a small mass ratio allows us to approach the contribution of the secondary object to the binary system in a perturbative way, i.e. to treat the secondary as a perturbation to a given black hole background. In particular, by expanding the background metric in terms of the mass ratio, we can calculate the gravitational self-force \cite{Barack19,Pound21}. The respective radiation reaction
carries away from the binary energy and angular momentum in the form of gravitational waves causing the secondary to inspiral towards the primary.

The aforementioned dissipation due to radiation reaction is actually slow when compared to the orbital motion of the secondary around the primary. This allows us to use a two timescale approach to model an EMRI \cite{Flanagan,kevorkian2012multiple}. The slow time scale is concerned with the evolution of the constants of motion, which correspond to the action variables of the system, while the fast time scale is concerned with the orbital phases (or ``orbital anomalies'') of the secondary, which correspond to the angle variables of the system \cite{arnold2007mathematical}. Hence, expressing an EMR system in action angle variables is a natural way to capture its dynamics. 

In Ref.~\cite{Schmidt02}, following the action-angle line of thought, Schmidt was able to compute the fundamental orbital frequencies of a body moving on geodesics in a Kerr black hole background. Ref.~\cite{Drasco04} went one step further the above work in the direction of EMRIs, when the authors used the fundamental frequencies to efficiently decompose the Teukolsky equation \cite{Teukolsky:1973ha} in the frequency domain in order to find the energy and angular momentum fluxes emitted by the secondary. Many other works employed the idea that the system describing an EMR should be, in principle, reexpressed in action-angle variables  \cite{Flanagan,Maarten14,LeTiec:2011ab,LeTiec:2015kgg, Fujita:2016igj,Isoyama:2018sib}. Refs. \cite{Fujita:2009bp,vandeMeent:2019cam} derived integral and special-function formulas for the transformation to action-angle coordinates for bound geodesics in Kerr space-time, but no work expressed the full \textit{closed-form} transform to \textit{and from} action-angle coordinates for black hole geodesics. This was only achieved in Ref.~\cite{Witzany22} and the present work. In Ref.~\cite{Witzany22} a Taylor series like approach has been used to find the action-angle variables for bound geodesics as a function of the energy and the angular momentum on a Schwarzschild background, while in the present work we employ a Lie series approach based on canonical perturbation theory (see, e.g., \cite{Efthymiopoulos11,cary1981lie} for a comprehensive introduction into this theory).

Canonical perturbation theory has a long history of successes,  such as the celebrated Kolmogorov-Arnold-Moser (KAM) theorem \cite{Arnold63} or the computation of asymptotic manifolds \cite{Moser58} (for more see, e.g., \cite{Contopoulos02,Morbidelli02}). Our study uses the framework of this theory to address the problem of EMRIs in a background dominated by a Schwarzschild black hole and perturbed by a surrounding matter field. The Lie series approach simplifies the system and allows us to have the respective Hamiltonian expressed purely in terms of actions. This implies that we have all the important quantities, such as the characteristic frequencies of motion in closed form. The Lie series approach also provides a canonical transformation to the action-angle variables, which implies that we have at hand the invertible mapping between the original coordinates and the action-angle ones. 

The perturbed black hole field that we study here can serve as a model for a broad range of physical scenarios. Massive black holes in the centers of galaxies are well known to be surrounded by dense nuclear star clusters and other molecular and dust structures \cite{neumayer2020nuclear,genzel2010galactic}. Other possibilities of external gravitational perturbations come from more exotic sources such as dark matter \cite{Hannuksela20,Macedo13} or scalar fields \cite{Ferreira17}. On the other hand, various physical effects such as tidal forces, increasing rotational shearing and the associated instabilities, or gravitational radiation lead to the evacuation of the immediate vicinity of the black hole with only a few massive objects remaining in the inner few hundred Schwarzschild radii \citep{merritt2013dynamics}. This motivates our approach, where the external matter distribution is considered as far from the black hole and its field expanded only to a handful of leading external multipoles. 

To date, the motion in the fields of black holes with external gravitational perturbations was mostly studied through the methods of numerical integration. Refs \citep{Vieira1996,deMoura2000} showed that the motion of free test particles in black hole fields superposed with external multipoles (exactly as we study here) corresponds to a weakly non-integrable system with the appearance of resonances and chaos. Several other works have since documented these properties for various exact matter sources outside of the black hole and by using a number of methods of numerical analysis \cite{vieira1999relativistic,semerak2010free,semerak2012free,Sukova13,witzany2015free,polcar2019free,polcar2019melnik}. Our work stands out by using an analytical method, while we have to be aware of its limitations due to the weak non-integrability of the system proven by the large body of numerical studies.

All the advantages of using the Lie series approach come with the cost that the new system is faithful to the original one only up to a certain accuracy. Nevertheless, every model has such flaws, the true issue is whether the approximation used is accurate enough for the purpose it will be used for. In this sense, this work provides a proof of principle that by using the Lie series approach one is able to compute fast adiabatic EMRIs with fair accuracy even in the case of more complex black hole backgrounds.

The rest of the article is organized as follows. Sec.~\ref{sec:pert} briefly introduces the perturbation theory method. Sec.~\ref{sec:circ} describes the background on which the inspirals evolve in our study, and details the methodology we followed to describe the geodesic motion on this background using the Lie series method. Sec.~\ref{sec:insp} discusses the techniques used to generate the inspirals, while Sec.~\ref{sec:results} presents our numerical results. Finally, Sec.~\ref{sec:Concl} discusses our main conclusions, and refinements and further steps that will be needed for applications of this approach in contexts such as the production of waveforms for LISA.

\section{Canonical perturbation theory} \label{sec:pert}

In this section we summarize the general method that will be used to cast the conservative dynamics in action-angle coordinates in Section \ref{sec:circ}. 

\subsection{Action-angle coordinates}
Consider a Hamiltonian system of $N$ degrees of freedom with the Hamiltonian $H(q_i,p_i)$ satisfying the following conditions:
\begin{enumerate}
  \item The system possesses $N$ linearly independent isolating integrals of motion   $I_i,\enspace \lbrace I_i, I_j\rbrace = 0,\enspace  i,j=1,\ldots N $
  \item Motion in the phase space is bounded.
\end{enumerate}
According to the Liouville-Arnold theorem (see for example \citep{Wiggins,arnold2007mathematical}), the motion is then confined to an $N$-dimensional subset of the phase space diffeomorphic to the torus $\mathbb{T}^N$. The particular torus on which the selected trajectory lies is defined by $N$ parameters $J_i$ called actions and can be parametrized by $N$ periodic angles $\psi_i\in(0,2\pi)$. Angles are canonically conjugate to actions and together they are known as action-angle coordinates.

If we perform a canonical transformation to action-angle coordinates we find that the new form of the Hamiltonian has one remarkable property, it depends only on the actions
\begin{align*}
   (q_i,p_i)\rightarrow (\psi_i,J_i),\enspace H(q_i,p_i)\rightarrow H(J_i) \,.
\end{align*}
Consequently the solution to Hamilton equations is trivial
\begin{align}\label{hameq}
\begin{aligned}
J_i(t)=\mathrm{constant},  \\
\psi_i(t)=\Omega_i t+\psi_{0i},
\end{aligned}
\end{align}
where $ \Omega_i=\frac{\partial H(J_j)}{\partial J_i} $ are the frequencies of motion. Inserting the solution \eqref{hameq} into the transformation relations we obtain
\begin{align}\label{transform}
q_i=q_i(\psi_j(t),J_j),\enspace p_i=p_i(\psi_j(t),J_j)\,.
\end{align}
Thus, finding the transformation from action-angle coordinates to the original coordinates is equivalent to solving the Hamilton equations of motion in the original coordinates $(q_i, p_i)$.

In the particular case of a separable system, in which the motion is periodic in coordinate $q_i$, the corresponding action can be computed using the formula \citep{Lichtenberg}
\begin{align}\label{action}
J_i=\frac{1}{2\pi}\oint p_i \mathrm d q_i\,,
\end{align}
where the integral is taken along a complete time period of the motion.

Unfortunately, there are very few examples in which the integral \eqref{action} can be evaluated in a closed form. One such example is the harmonic oscillator whose Hamiltonian can be transformed as
\begin{align}\label{oscillator}
H(q,p)=\frac{p^2}{2m}+\frac{1}{2}m\Omega^2q^2 \rightarrow \enspace H(\psi,J)=\Omega J \,,
\end{align}
while the transformation relations are given by
\begin{align}
 q=\sqrt {{\frac {2J}{m\Omega}}}\sin \left( \psi \right),\hspace{35pt} p=\sqrt {2J\Omega\,m}\cos      \left( \psi \right). 
\end{align}
In these relations one can clearly see the familiar solution to the harmonic oscillator problem.

At this point one can surely ask whether it is possible to perform such a transformation in case of a more complicated integrable system or even in a case of a slightly perturbed integrable system. This question leads us directly to the Lie series formalism.
 
\subsection{Lie series} \label{sec:LieS}

The Lie series are a class of canonical transformations defined by an arbitrary generating function $\omega(q_i,p_i)$ (details in Refs \citep{Efthymiopoulos11,cary1981lie,Deprit}). We will first describe this on the best known case of time evolution. Denoting $z_i$ as the phase space coordinates, the evolution equation can be written as
 \begin{align}\label{evolution}
\frac{\mathrm{d}z_i}{\mathrm{d}t}=\lbrace z_i,H\rbrace.
\end{align}
The solution to this equation can then be found using the Taylor expansion where the time derivative is replaced by the Poisson bracket with the Hamiltonian, which is a consequence of eq. \eqref{evolution}. Denoting $z_i(0)=z_i$ we have
\begin{align}
\begin{split} 
z_i(t)=z_i+\lbrace z_i,H\rbrace t+\frac{1}{2} \lbrace \lbrace z_i,H\rbrace,H\rbrace t^2+\ldots=\\ =\exp(t\pounds_{H}) z_i \,,
\end{split} 
\end{align}
where the operator $\pounds_g$ is defined as $\pounds_{g} f=\lbrace f ,g\rbrace$. The operator $\pounds_g $ is called Lie derivative because in the geometrical formulation of Hamiltonian mechanics we have $\pounds_g = \pounds_{X_g} $ where $X_g$ is the Hamiltonian vector field associated with the function $g$.

If we now make an exchange:
$H\leftrightarrow \omega(q_i,p_i),\enspace t \leftrightarrow \varepsilon  $ where $\varepsilon$ is a small parameter, we can define a new transformation
\begin{align}
Z_i=z_i(\varepsilon)=\exp(\varepsilon \pounds_{\omega})z_i\,.
\end{align}
This transformation is indeed canonical as the Poisson brackets are preserved due to the following identity \citep{Deprit}:
\begin{align}
\lbrace \exp(\pounds_{\omega})f, \exp(\pounds_{\omega})g \rbrace =\exp(\pounds_{\omega})\lbrace  f,g \rbrace,
\end{align}
which leads to $\lbrace z_i ,z_j\rbrace=\lbrace Z_i ,Z_j\rbrace$. Another useful identity is the inverse relation for the Lie operator
\begin{align}
\big(\exp(\varepsilon \pounds_{\omega})\big)^{-1}=\exp(-\varepsilon \pounds_{\omega}).
\end{align}

Having introduced the Lie series formalism, we can now use it to approximately transform a Hamiltonian into action-angle coordinates, or to be more specific, to find the so called Birkhoff normal form of a Hamiltonian. 

  \subsection{Birkhoff normal form}
  
Assume we have a Hamiltonian in the form:
\begin{align}
 H^{(0)}=H_0(J_i)+\displaystyle\sum_{j=1}\varepsilon^j H_{j}^{(0)}(\psi_i,J_i) \,,
\end{align}
where $H_0(J_i)$ is a well known integrable Hamiltonian already expressed in the action-angle form while the other part is expanded in a small perturbation parameter $\varepsilon$.
Computing the Birkhoff normal form of $H^{(0)}$ actually means eliminating the angle variables from the Hamiltonian.

Starting from the first order of the perturbation we decompose the Hamiltonian $H_{1}^{(0)}$ into the part which does not depend on the angles and the other which does:  $H_{1}^{(0)}=Z_1(J_i)+h_1(\psi_i,J_i)$. If we then act with the Lie operator on $H^{(0)}$ we get
\begin{align}
 \exp(\varepsilon \pounds_{\omega_1})H^{(0)}=H_0+\varepsilon Z_1+\varepsilon\lbrace H_0,\omega_1\rbrace+\varepsilon h_{1}+\mathcal{O}(\varepsilon^2).
\end{align}
Since  $h_1$ is to be eliminated, the terms proportional to $\varepsilon$ have to satisfy:
\begin{align}\label{eq:HomoEq}
\enspace \lbrace H_0,\omega_1\rbrace+h_{1}\stackrel{!}{=}0.
\end{align}
This is called  homological equation which has to be solved for the so far unknown generating function $\omega_1$. Once we have found  $\omega_1$ we can compute a new form of our Hamiltonian
 \begin{align}
H^{(1)}=\exp(\varepsilon \pounds_{\omega_1})H^{(0)}=H_0(J_i)+\varepsilon Z_1(J_i)+\mathcal{O}(\varepsilon^{2}).
\end{align}
Thus, we have our Hamiltonian in the action-angle variables up to the first order in $\varepsilon$.
We can now proceed in a similar fashion, i.e., solving another homological equations and so on until arriving at a desired order $n$ in which the Hamiltonian reads 
  \begin{align}
  \begin{split}
H^{(n)}=\exp(\varepsilon^{n} \pounds_{\omega_{n}})\exp(\varepsilon^{n-1} \pounds_{\omega_{n-1}})\ldots \exp(\varepsilon \pounds_{\omega_{1}}) H^{(0)}=\\=U(\omega_i)H^{(0)},
 \end{split}
\end{align}
where the notation $U(\omega_i)$ is used just for brevity to represent the $n$ canonical transformations applied. Furthermore, $H^{(n)}$ can be decomposed into
  \begin{align}
  \begin{gathered}
  H^{(n)}=H_{NF}(J_i)+R^{(n)}(\psi_i,J_i) \,,\\
H_{NF}(J_i)=H_0(J_i)+\displaystyle\sum_{j=1}^{r}\varepsilon^j Z_{j}(J_i);\enspace  R^{(n)}(\psi_i,J_i)=\mathcal{O}(\varepsilon^{n+1}) ,
\end{gathered}
\end{align}
where $H_{NF}$ is the Birkhoff normal form of $n$th order, while $R^{(n)}$ is a remainder which can be neglected as $\varepsilon^{n+1}$ is a sufficiently small number.   
The Birkhoff normal form now allows us to compute the frequencies of motion which tell us how the new angles evolve (see Eq.~\eqref{hameq}). The old coordinates can now be expressed  in terms of the new ones
 \begin{align}
\psi^{(0)}=U(\omega_i)\psi,\enspace J^{(0)}=U(\omega_i)J\,.
\end{align}
Inserting Eq.~\eqref{hameq} into the transformation relation \eqref{transform} gives us an approximative solution to the equation of motion with the error given by the size of the remainder $R^{(n)}$.

The Lie series is generally only asymptotic; there may exist a maximum order above which the approximation becomes less and less precise. The question of convergence of canonical perturbation theory is tied to the existence of small divisors and resonances, and the Lie series is not guaranteed to converge everywhere even in fully integrable systems  (see, e.g., Refs \citep{Efthymiopoulos11,arnold2007mathematical}). We shall demonstrate the issues with resonances also in Sec.~\ref{sec:validity}.

\section{Tidally perturbed black hole orbits} \label{sec:circ}

Here we compute the conservative evolution of mildly eccentric orbits of test particles near black holes perturbed by a faraway gravitating ring surrounding the system. We first introduce the metric field and then apply the Lie series method to obtain action-angle coordinates of near-circular geodesics in this field. Consequently, we apply another round of canonical perturbation theory to obtain the approximate solution of these orbits under the tidal perturbation by the ring. This will be a basis for the adiabatic inspirals computed in Sec.~\ref{sec:insp}.

\subsection{A black hole perturbed by a ring-like source}

Picture a black hole of mass $M$ encircled by a rotating gravitating ring with mass $\mathcal{M}_{\rm r}$ and radius $r_{\rm r}\gg M$ much larger than the black hole horizon. What are going to be the leading-order effects of the ring on the gravitational field near the black hole? It was found already by Thirring in 1918 \citep{thirring1918wirkung,Mashhoon:1984fj} that the local inertial system inside a light, thin rotating shell is rotating with respect to the inertial system at infinity with an angular velocity
\begin{align}
    \Omega_{\rm Thir} = \frac{2 \mathcal{J}_{\rm sh}}{ r_{\rm sh}^3}\,,
\end{align}
where $\mathcal{J}_{\rm sh}, r_{\rm sh}$ are the total angular momentum and radius of the shell. Similarly, the rate of time inside the shell is redshifted with respect to observers at infinity by the gravitational potential on the surface of the shell
\begin{align}
    z_{\rm sh} = \frac{\mathcal{M}_{\rm sh}}{r_{\rm sh}} \,,
\end{align}
where $\mathcal{M}_{\rm sh}$ is the total mass of the shell.

Based on the works of Refs \cite{Will:1974zz,Cizek:2017wzr}, we show in Appendix~\ref{app:ring} that a similar effect occurs in the case of the ring-hole system. Specifically, the inertial frame near the center of a ring of angular momentum $\mathcal{J}_{\rm r}$ and Schwarzschild radius $r_{\rm r}$ rotates, to leading order in $r_{\rm r}\gg M\gg \mathcal{M_{\rm r}}$, with an angular velocity
\begin{align}
    \Omega_{\rm in} = \frac{2\mathcal{J}_{\rm r}}{r_{\rm r}^3} +\mathcal{O}(r_{\rm r}^{-5})\,.
\end{align}
We assume that the ring-like structure is moving approximately as a test body in the black-hole field, so to leading order we have $\mathcal{J}_{\rm r}  = \mathcal{M}_{\rm r}\sqrt{M r_{\rm r}^2/(r_{\rm r}-3M)}$ and we can write
\begin{align}
    \Omega_{\rm in} = 2\mathcal{M}_{\rm r} \sqrt{\frac{M}{r_{\rm r}^5}}\left(1 + \frac{3 M}{2 r_{\rm r}}\right) + \mathcal{O}(r_{\rm r}^{-9/2})\,.
\end{align}
Additionally, the internal frame is redshifted by a factor
\begin{align}\label{redshift}
    z_{\rm in} = \frac{\mathcal{M}_{\rm r}}{r_{\rm r}}\left(1 + \frac{M}{r_{\rm r}}\right) + \mathcal{O}(r_{\rm r}^{-3})\,.
\end{align}
Finally, the gravitational field near the black hole will also have a tidal contribution from the ring. Apart from assuming that the black hole is static in the ``internal" inertial frame, we also truncate the tides to leading quadrupolar order. Then we obtain the metric valid near the black hole (see Appendix~\ref{app:ring} for details): 
\begin{align}
    \begin{split}
    & \d s^2_{r\ll r_{\rm r}}  = -\left(1 - \frac{2M}{r}\right)(1 + 2 \nu_{Q})\d t^2 + \frac{1 + 2 \chi_Q - 2 \nu_Q}{1 - 2M/r} \d r^2 \label{eq:metric} \\ 
    & \phantom{\d s^2  =}  + (1 - 2 \nu_Q)r^2 \left[(1 + 2\chi_Q) \d\th^2 + \sin^2\!\th\d\ph^2\right]\,,
    \end{split} \\
    & \nu_Q \equiv \frac{Q}{4} \left[r(2M-r)\sin^2\!\th + 2(M-r)^2 \cos^2\!\th-6M^2 \right] \label{eq:gravpot} \,, \\
    & \chi_Q \equiv QM(M-r) \sin^2\!\th \,,
\end{align}
where $Q \equiv \mathcal{M}_{\rm r}/r_{\rm r}^3$ is the quadrupole perturbation parameter and $t,r,\th,\phi$ are Schwarzschild-like coordinates in the local frame. The local metric is approximately vacuum, static and axisymmetric with respect to the local time and azimuthal angle $t,\ph$ with corresponding Killing vectors $\xi^\mu_{(t)} = \delta^\mu_t,\, \xi^\mu_{(\ph)} = \delta^\mu_\ph$. As such, it is an approximate Weyl metric \citep{weyl1917gravitationstheorie,Griffiths:2009dfa}.

The form \eqref{eq:metric} of the metric is valid only for $r\ll r_{\rm r}$ and rings that are not compact, $\mathcal{M}_{\rm r} \ll r_{\rm r}$. Specifically, it neglects all terms starting from $\mathcal{O}(r_{\rm r}^{-4})$ and $\mathcal{O}(\mathcal{M}_{\rm r}^2)$. Also, as already discussed, the local coordinates are related to the coordinate time and azimuthal angle $T, \phi$ of static observers at infinity as
\begin{align}
    & \d T = (1 + z_{\rm in}) \d t\,, \\
    & \d \phi = \d \ph + \Omega_{\rm in} \d t \,.
\end{align} 
This has to be taken into account when predicting observations from the dynamics in the metric field \eqref{eq:metric}.

Note that this framework is very flexible, since it does not necessarily fix the relationship between $z_{\rm in}, \Omega_{\rm in}$ and $Q$. For more general matter distributions than a thin ring, these parameters can be computed separately and fed into the formalism the same way as it is done here. However, one case which we do not treat are time-dependent and non-axisymmetric perturbations which would correspond to perturbers orbiting our EMR binary at intermediate distances. The possibility of inclusion of this case is discussed in Sec.~\ref{sec:Concl}.

\subsection{Quasi-circular Schwarzschild geodesics}

We would now like to apply the aforementioned theory to the Hamiltonian 
\begin{align}\label{eq:ham}
H_{\rm tot}=\frac{1}{2}g^{\mu\nu}p_\mu p_\nu\, ,
\end{align}
where $g^{\mu\nu}$ is our background metric \eqref{eq:metric} and the four-momentum $p_\mu$ is normalized to unity, i.e. $g^{\mu\nu}p_\mu p_\nu=-1$. We will first start with the well known Schwarzschild Hamiltonian ($Q=0$)
\begin{align}\label{eq:HamSchw}
H_{\rm Schw}=\frac{1}{2}\left[\frac{-1}{1-\frac{2M}{r}} p_t^2+\bigg(1-\frac{2M}{r}\bigg)p_r^2+\frac{1}{r^2}\left(p_\theta^2+\frac{p_\phi^2}{\sin^2\theta}\right)\right].
\end{align}
Unfortunately this Hamiltonian cannot be put exactly into the action-angle variables like the Kepler Hamiltonian. Nevertheless  $H_{\rm Schw}$ remains separable which means  that by adopting a new evolution parameter $\lambda$ defined as $\mathrm{d}\tau=r^2\mathrm{d}\lambda$ we can separate our Hamiltonian to a radial and an angular part.  The parameter $\lambda$ is a special case of the Carter-Mino time \cite{Carter:1968rr,Mino} used for  similar reasons in the Kerr spacetime. The Hamiltonian generating evolution in $\lambda$  (see Appendix~\ref{app:Schw}) can be written as 
\begin{align}\label{eq:AAIsol4}
H_{{ \rm Schw}(\lambda)}=\frac{1}{2}r^2 (g_{S}^{\mu\nu}p_\mu p_\nu+1)= H_{\rm rad}+H_{\rm ang} \, ,
\end{align}
where 
 \begin{align}\label{eq:Hrad}
  H_{\rm rad} &= \frac{1}{2}r^2\left[-\frac{1}{1-\frac{2M}{r}} p_t^2+\bigg(1-\frac{2M}{r}\bigg)p_r^2+1\right] \, ,
 \end{align}
is the radial part and $$H_{\rm ang}=\frac{1}{2} \left( p_\theta^2+\frac{p_\phi^2}{\sin^2\theta}\right)$$ is the angular part of the Hamiltonian.

Having separated our Hamiltonian we can now perform the transformation of the respective parts into action-angle coordinates. Since the metric does not depend on $\phi$, the specific angular momentum  $p_\phi=J_\phi=L_z$, i.e. the $z$-component of the angular momentum per unit mass, is already an action variable while for the $\theta$ part we have
\begin{equation}\label{eq:AAIsol5}
J_\theta=\frac{1}{2\pi}\oint p_\theta \mathrm d \theta =L-J_\phi
\end{equation}
where $L$ is the specific total angular momentum. The action $J_\theta$, thus, describes the part of angular momentum associated with non-equatorial motion.\footnote{For equatorial motion  $J_\theta=0$.} The angular part of the Hamiltonian in the action-angle coordinates reads
\begin{equation}\label{eq:angularH}
H_{\rm ang}=\frac{1}{2}(J_\theta+J_\phi)^2 \, .
\end{equation}
The conjugate angles $\psi_\theta$ and $\psi_\phi$ to the actions $J_\theta$ and $J_\phi$ are obtained by canonical transformations, which can be found in Appendix~\ref{app:Schw}.

Let us now discuss the more difficult part, which involves the radial part $H_{\rm rad}$. In Eq.~\eqref{eq:Hrad} we replace the $p_t$ component of the four-momentum by the specific energy of the system $E=-p_t$ and we perform an expansion of the system around a stable circular orbit ($r=r_c, p_r=0, E=E_c$). The perturbation parameter along which the expansion takes place is the distance from the circular orbit (something akin to the eccentricity). We assume that the relevant phase space coordinates deviate from those corresponding to the circular orbit like 
\begin{align}\label{eq:expansion}
   r-r_\mathrm{c}=\mathcal{O}(\varepsilon)=p_r,\enspace  E-E_c=\delta E=\mathcal{O}(\varepsilon^{2})\, ,
\end{align}
where $\varepsilon$ is a book-keeping parameter telling us how big each term in our expansion is. Keep in mind that $\varepsilon$ is not the perturbation parameter, after the computation we can just set it to $\varepsilon=1$.

As the radius  $r_\mathrm{c}$ is the minimum  of the effective potential $V_{\rm eff}=H_{\rm rad}$ (see, e.g., \citep{Chandrasekhar:1985kt}), the first post-circular approximation is the harmonic oscillator.  After performing a transformation similar to Eq.~\eqref{oscillator} (details in Appendix \ref{app:Schw}) we get the radial part in the form
 \begin{align}
  H_{\rm rad} &= K_0+K_2\delta E+J_r\Omega_{rc} \nonumber +R(\delta E,\psi_r,J_r)\, ,
 \end{align}
where $K_0$ and $K_2$ are constants and $\Omega_{rc}$ is the frequency of the respective harmonic oscillator (Appendix \ref{app:Schw}).
 
Neglecting the remainder $R(\delta E,\psi_r,J_r) = \mathcal{O}(\varepsilon^3)$ we could accurately describe quasi-circular orbits. However we would like to describe nearly all the bound orbits with sufficient precision, which is why we implement the canonical perturbation theory as discussed in the next section.


\subsection{Tidally perturbed orbits}

In this section we finish the construction of the approximative Hamiltonian system of a Schwarzschild with a ring in action-angle variables. First, we perform two normalization steps, as described in Sec.~\ref{sec:LieS}. This implies finding two generating functions $\omega_{1}$ and $\omega_{2}$ to be used in the Lie operators acting on the Hamiltonian $H_{{\rm Schw}(\lambda)}$
\begin{equation}\label{eq:2OHschw}
 \exp( \pounds_{\omega_{2}})\exp( \pounds_{\omega_{1}})H_{{\rm Schw}(\lambda)}=H_{\rm NS}(J_r,J_\theta)+\mathcal{O}(\varepsilon^{5}) \, .
\end{equation}
For the purposes of our study we deem approximation~\eqref{eq:2OHschw} to be sufficiently describing geodesic bound orbits around a Schwarzschild black hole, hence, we can now add the ring-like source.

Our initial Hamiltonian~\eqref{eq:ham} can be naturally split into the Schwarzschild and the ring part as in the case of the linearly perturbed metric~\eqref{eq:metric}
\begin{equation}\label{eq:Htot}
 H_{\rm tot}=H_{{\rm Schw}(\lambda)}+Q H_{\rm ring}\, .
\end{equation}
The perturbation part is then transformed into the same coordinates as the Schwarzschild part
\begin{equation}\label{eq:2OHtot}
\exp( \pounds_{\omega_{2}})\exp( \pounds_{\omega_{1}}) H_{\rm ring}=H_{Q1}+\mathcal{O}(\varepsilon^{5})\, ,
\end{equation} 
while the terms of higher order in $\varepsilon$ are neglected. And, thus, the total Hamiltonian reads:
 \begin{equation}\label{eq:Htot1}
H_{\rm tot}=H_{\rm NS}(J_r,J_\theta)+Q H_{Q1}(\psi_r,\psi_\theta,J_r,J_\theta)+\mathcal{O}(\varepsilon^{5})\, .
\end{equation} 
The last step in our computation is to solve the homological equation for the function $\chi$ in order to eliminate the angles from $H_{Q1}$.\footnote{Note that now $Q$ is the perturbation parameter.} After this, the total Hamiltonian reads
\begin{align}\label{eq:HamNF}
 H_{\rm tot}&=H_{\rm NS}(J_r,J_\theta)+Q Z_{Q1}(J_r,J_\theta)+\mathcal{O}(Q^{2}) \nonumber \\
  &=H_{\rm N}(J_r,J_\theta)+\mathcal{O}(Q^{2}) \, .
\end{align} 

The original coordinates can be expressed using the Lie operators as functions of the new ones as follows:
\begin{align}\label{eq:nc2oc}
 r &=\exp(Q\pounds_{\chi})\exp( \pounds_{\omega_{2}})\exp( \pounds_{\omega_{1}})r_{0}=U(\omega_{1},\omega_{2},\chi) r_{0}, \nonumber \\ 
 p_r &=U(\omega_{1},\omega_{2},\chi) p_{r_{0}},  \\
 \theta &=\exp(Q\pounds_{\chi}) \theta_{0},\; p_\theta=\exp(Q\pounds_{\chi}) p_{\theta_{0}}, \; \phi=\exp(Q\pounds_{\chi}) \phi_{0} \nonumber
\end{align}
where $r_{0}, p_{r_{0}}, \theta_{0}, p_{\theta_{0}}$  and $\phi_{0}$ are the original transformation functions given by Eqs.~\eqref{eq:AAIsol9}, \eqref{eq:thvar2aa},  and \eqref{eq:phi} respectively in the Appendix~\ref{app:Schw}. 

Solving the homological equation~\eqref{eq:HomoEq} at each step is quite straightforward, the relatively difficult part is finding the generating function $\chi$ as it involves two degrees of freedom. By expanding the Hamiltonians in parameters $\varepsilon$ and $Q$, the $h_1$ part of $H_{Q1}$, which is to be eliminated, takes form 
 \begin{equation}\label{eq:hom1f}
h_{1}=\displaystyle\sum_{k,l} a_{kl} e^{i(k\psi_r+l\psi_\theta)},
\end{equation}
where the coefficients $a_{kl}$ are in fact functions of actions. The solution to Eq.~\eqref{eq:HomoEq} can then be expressed as:
\begin{equation}\label{eq:gf_exp}
 \chi=\displaystyle\sum_{k,l} a_{kl} \frac{1}{i(k\Omega_{r0}+l\Omega_{\theta0})}e^{i(k\psi_r+l\psi_\theta)}
\end{equation}
where $\Omega_{r0}=\frac{\partial H_{\rm NS}}{\partial J_r} $ and $\Omega_{\theta0}=\frac{\partial H_{\rm NS}}{\partial J_\theta}$ are the frequencies of the Schwarzschild Hamiltonian obtained in \eqref{eq:2OHschw}. When close to resonances the denominator of this expression tends to zero which causes the remainder to be large, thus making the approximation less accurate as we shall see in the following section. 

All the above-mentioned calculations involving canonical perturbation theory are included in the first part of our Supplemental material \cite{SupMat1}.

\begin{figure}[h!] \centering
	\includegraphics[width=0.95\linewidth]{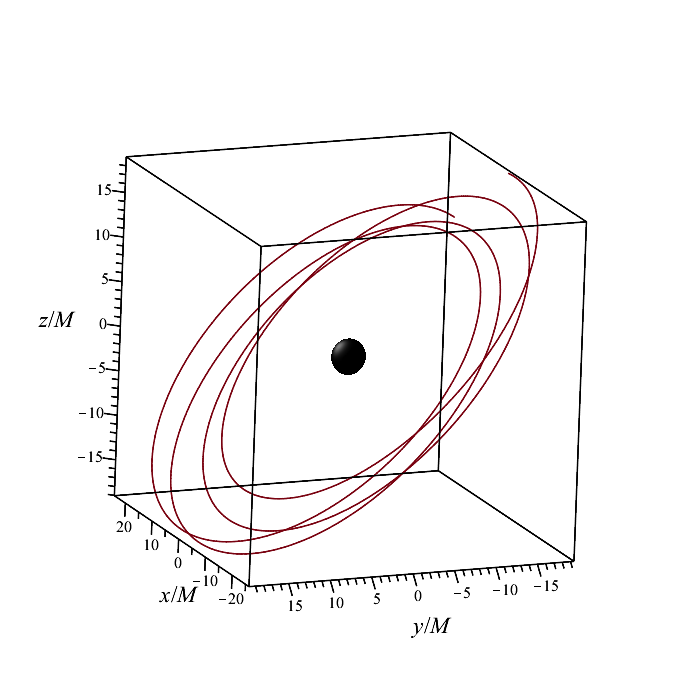}
\caption{Spatial representation of the approximate solution to the geodesic equation. ($Q=10^{-6} M^{-2}, J_r=0.1M,  J_\theta=1.5 M, J_\phi=3.5M$.) } \label{fig:3dplot}
\end{figure}

The analytical formulas~\eqref{eq:nc2oc} can  be plotted for fixed values of the actions to illustrate our result (Fig.~\ref{fig:3dplot}). It is clear from the figure that nonequatorial motion is no longer planar since the ring (located in the $z=0$ plane) breaks the spherical symmetry of the Schwarzschild spacetime.

\subsection{Validity of the approximation} \label{sec:validity}

Knowing the explicit expressions \eqref{eq:nc2oc} and the approximate  normal form of our Hamiltonian $H_{\rm N}(J_i)$ \eqref{eq:HamNF} we have essentialy perturbatively solved the original equations of motion given by $H_{\rm tot}$.  First we fix three of our new set of conserved actions and then compute the fourth so that the normalization condition 
\begin{align}\label{eq:Normalization}
   H_N(J_r,J_\theta,\delta E,J_\phi)=0 
\end{align}
is satisfied. Then, we find the frequencies of motion for our new angles as well as the relation between the coordinate time $t$  and  $\lambda$
\begin{align}\label{eq:NewFr}
\frac{\mathrm{d}\psi_i}{\mathrm{d}\lambda}=\frac{\partial H_N}{\partial J_i}=\Omega_i(J_r,J_\theta,\delta E,J_\phi), \hspace{25pt}
\frac{\mathrm{d}t}{\mathrm{d}\lambda}=\frac{\partial H_N}{\partial \delta E}.
\end{align}
Finally we substitute the angles and actions into Eq.~\eqref{eq:nc2oc} to get the coordinates and their respective momenta  as explicit functions of Mino time $\lambda$ $(x^i(\lambda),p_i(\lambda))$.

The evolution of the deviations from the exact solutions is governed by the remainder $R(\psi_i,J_i)$ which contains all the terms of the order $\mathcal{O}(\varepsilon^{5})$  and $\mathcal{O}(Q^{2})$. It is clear that the validity of our approximation not only depends on the fixed parameter $Q$ describing the gravitational field of our ring-like source, but also on all our actions $J_r, J_\theta, J_\phi$. The most straightforward way to test our approximation would be plotting and comparing our analytical solution to the numerical one. This is certainly illustrating, nevertheless, it is still useful to have some quantity to describe the deviation from the exact solution. For this purpose we can use quantities denoted as $\delta J_i$ which measures the relative error of the conservations of actions. The errors $\delta J_i$ can then be expressed as functions of actions  $J_i$ in order to study the validity of the approximation (details are given in Appendix~\ref{app:tests}). 

In general, it can be said that the larger the value of $J_r$ is the less accurate the approximation becomes. Apart from that the approximation depends on the perturbation parameter $Q$ and the total angular momentum $L$. These two, however, are not independent from each other
as increasing the value of $L$ is equivalent to increasing $Q$. This comes from the fact that $\nu_Q$ is not bounded by a fixed value of $Q$ instead we have $\nu_Q \sim r^2$ (see Eq.~\eqref{eq:gravpot}) and $r\sim L^2$. We can, thus, conclude that it is the value of the quantity $Q L^4/M^2$ that characterizes the entire strength  of the perturbation. 

\begin{figure}[h!] \centering
	\includegraphics[width=0.85\linewidth]{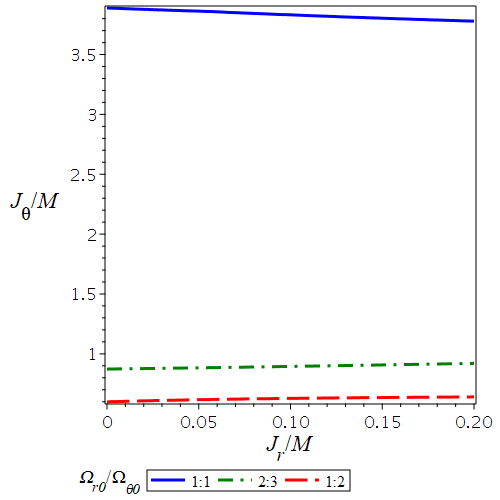}
\caption{Resonances in the  $J_r$-$J_\theta$ plane for $J_\phi=3M$, $Q=10^{-5} M^{-2}$.} \label{fig:resoloc}
\end{figure}

The most general type of motion is the nonequatorial one, for which we have $J_\theta \neq 0 $. The dependence of $\delta J_r$ and  $\delta J_\theta$ on actions  is the same as in the equatorial motion, what is new here, however, is the presence of the resonances of the form 
\begin{align}\label{eq:reson}
k  \Omega_{r0}+l  \Omega_{\theta0}=0 , \enspace   k, l \in \mathbb{Z}\setminus\lbrace 0\rbrace \, ,
\end{align}
where $\Omega_{r0}$ and $\Omega_{\theta0}$ are the Schwarzschild frequencies. When the orbit is close to these resonances the approximation is no longer reliable because the denominators in the expression~\eqref{eq:gf_exp} tend to zero. These small divisors then prevent the convergence of the normalization procedure (see, e.g., \cite{Efthymiopoulos11}). In our particular case we only applied one generating function involving two degrees of freedom which is $\chi$ (used in Eq.~\eqref{eq:HamNF}). From the analysis of the function $\chi$ and the Schwarzschild frequencies it becomes clear that the only  ratios $\frac{\Omega_{r0}}{\Omega_{\theta0}}$ present in the expansion \eqref{eq:gf_exp} of $\chi$ are the $\frac{1}{2}$, $\frac{2}{3}$ and $1$. This is illustrated in Fig.~\ref{fig:resoloc}, which depicts the resonance sets in the  $J_r$-$J_\theta$ plane for a fixed value of  $J_\phi$. It is important to stress at this point that even though $\Omega_{r0}$  depends explicitly on the total angular momentum $L=J_\theta+J_\phi$ it is correct to treat both angular actions separately. The same can be said for the dependence on $Q$, which is much more significant than that on $L$, e.g. smaller value of $Q$ shifts the resonance curves to larger values of $L$. This dependence can be understood from the substitution of the energy from constraint~\eqref{eq:Normalization} in $\Omega_{r0}$, for which we have $E=E(Q,J_r,J_\theta,J_\phi) \neq E(Q,J_r,J_\theta+J_\phi)$. Namely, in total the  expression for the radial frequency reads  $\Omega_{r0}=\Omega_{r0}(E(Q,J_r,J_\theta,J_\phi),J_r,J_\theta+J_\phi)$.

In summary, the  geodesics obtained from the Hamiltonian~\eqref{eq:HamNF} approximate the exact solution sufficiently well provided that the errors $\delta J_i$ are small and we are not close to resonances. However, the size of $\delta J_i$ unfortunately does not tell the whole story. The error in the evolution of $(\psi_i, J_i)$ accumulates over time and it is up to us to fix the accuracy of the approximation, so the error does not become substantial for $\varepsilon_m^{-1}$ number of orbital periods (Appendix~\ref{app:tests}), where $\varepsilon_m$ is the mass ratio. 

\begin{figure*}[htp!]
	\includegraphics[width=0.33\linewidth]{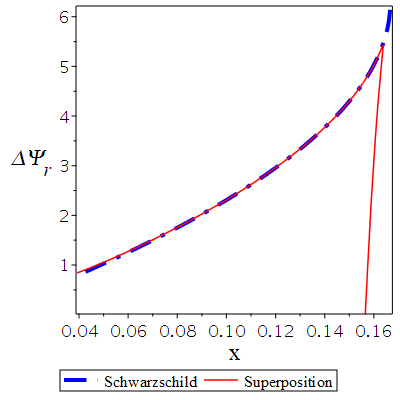}
	\includegraphics[width=0.33\linewidth]{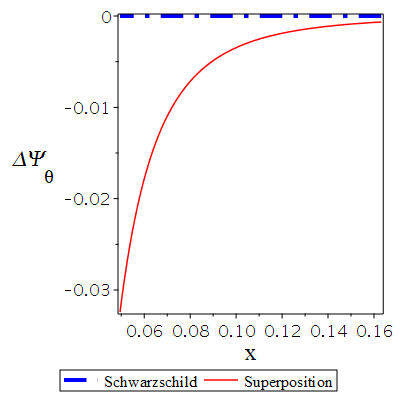}
	\includegraphics[width=0.33\linewidth]{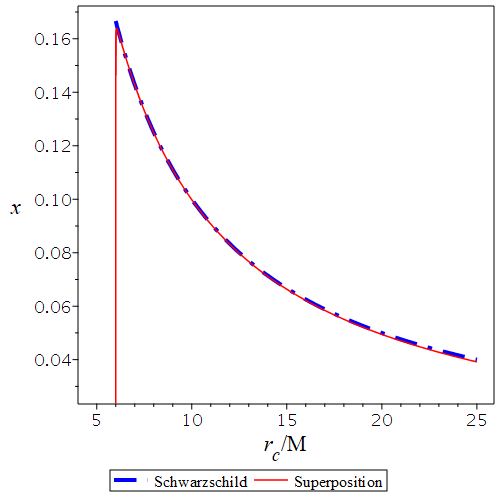}
\caption{The relation between  pericenter (left panel) and the nodal (middle panel) precession rate of near-circular near-equatorial orbits and a frequency parameter $x$ which itself can be expressed as a function of $r_c$ (right panel), $Q=10^{-6} M^{-2}$.}
\label{fig:prec}
\end{figure*}

To conclude this part we illustrate the advantages and the drawbacks of this approximation on the precession of near-circular and near-equatorial orbits. The precession rates of orbits can be expressed using fundamental frequencies as 
\begin{align}\label{eq:prec}
\Delta \Psi_r=2 \pi \bigg( 1-\frac{\Omega_r}{\Omega_\phi}\bigg)\bigg\vert_{r_{\rm circ.}}\!, \;  \Delta \Psi_\theta=2 \pi \bigg( 1-\frac{\Omega_\theta}{\Omega_\phi}\bigg)\bigg\vert_{r_{\rm circ.}} \!,
\end{align}
where $\Delta \Psi_r$ corresponds to the pericenter precesion per one period of the azimuthal coordinate $\phi$  while similarly for the nodal precession rate we have  defined the quantity $\Delta \Psi_\theta$. Of course the frequencies are functions of the integrals of motion (actions) and for a near-circular and near-equatorial orbits they need to be evaluated at $(\delta E=0, J_r=0, J_\theta=0)$. The precession  rates are then functions solely of $J_\phi$, which itself can be expressed as a function of the radial coordinate $r$ or rather the circular-orbit location $r_c$. For the Schwarzschild  spacetime  the precession rates~\eqref{eq:prec} reduce to simple results 
\begin{align}
\Delta \Psi_r\vert_{Q=0} = 2\,\pi\, \left( 1- \sqrt{1-\,{\frac {6M}{{\it r_c}}}} \right) , \hspace{15pt} \Delta \Psi_\theta\vert_{Q=0} = 0.
\end{align}

Alternatively instead of $r_c$ we can use a dimensionless frequency parameter $x$
\begin{align}
x= (M \Omega_\phi^{(t)})^{\frac{2}{3}}, \hspace{15pt} \Omega_\phi^{(t)} = \Omega_\phi \frac{\mathrm{d\lambda}}{\mathrm{d}t}. 
\end{align}
where the azimuthal frequency $\Omega_\phi^{(t)}$ is defined with respect to the coordinate time $t$ (and not the Mino time $\lambda$). For Schwarzschild the frequency parameter $x$ is related to $r_c$  by a simple formula $x=\frac{M}{r_c}$. In the superposition background, however, the quantity $x$ is not an injective  function of $r_c$, which can be seen in  Fig.~\ref{fig:prec} (right panel). The same figure also shows the relation between $x$ and the precession rate of near-circular near-equatorial orbits. The nodal precession rate is in general nonzero and tends to grow (in absolute value) with the distance from the black hole, which can be expected since we are approaching the external gravitating ring that breaks the spherical symmetry. This in turn means that it is small for large values of $x$.

The left panel in Fig.~\ref{fig:prec} shows an unexpected divergence of $\Delta \Psi_r$ close to the innermost stable circular orbit\footnote{Note that the ISCO is slightly shifted from the Schwarzschild value $r=6M$} (ISCO). This we deem to be completely unphysical as it happens in the region dominated by the black hole. The divergence is in fact caused by the factor $\Omega^{-1}_{rc}$ present in the generating functions $\omega_1$ and  $\omega_2$ as  $\Omega_{rc}=0$ for $r_c=6M$. The sudden decrease of $x$ at Schwarzschild ISCO is caused for the very same reason. One would be able to invert the function $x(r_c)$ if not for this unphysical part of the graph. The fact that  $x(r_c)$  cannot be inverted can be also seen from the graph of  $\Delta \Psi_r$. Actually, this effect is a consequence of the perturbation expansion we have chosen. 

Had we used the expansion from the circular equatorial orbits of the superposition, the unphysical part of the graph would not have appeared  as the fundamental frequencies of the circular orbits are finite. In fact, we have checked this claim numerically. On the other hand  the expansion scheme we have used is easier to perform due to the simplicity of the Schwarzschild Hamiltonian and it can also describe orbits with arbitrary inclination.  Nevertheless, note that close to the Schwarzschild ISCO the approximation fails to describe the correct geodesic dynamics anyway.

\section{Adiabatic inspirals into the perturbed black hole} \label{sec:insp}

In this section we are going to present our model of an EMRI in the perturbed background field by using a basic prescription for radiation reaction. In the geodesic context we defined our Hamiltonian~\eqref{eq:ham} using a four-momentum normalized to $-1$. To reintroduce the mass $m$ of our ``particle'', we retain the original form of our Hamiltonian with actions and energy normalized to unit mass, $E=E^{(m)}/m$, $J_i=J_i^{(m)}/m$ instead of using the new coordinates $J_i^{(m)}$ and normalization to $-m^2$. Regardless of that, every not dimensionless quantity is still scaled with respect $M$ as can be seen in all the figures presented in our paper.

\subsection{Computation of the gravitational-wave fluxes}

Following the general approach described in \cite{Flanagan,kevorkian2012multiple},  we can write down the equations of motion of an inspiraling binary as an expansion in the mass ratio $\varepsilon_m$
 \begin{align}
& \frac{\mathrm{d}\psi_i}{\mathrm{d}t}= \Omega_i^{(t)} (\mathbf{J})+\mathcal{O}(\varepsilon_m) \label{eq:evolutioneq1} \,,\\
& \frac{\mathrm{d}J_i}{\mathrm{d}t}=\varepsilon_m G_i(\mathbf{\psi},\mathbf{J})+\mathcal{O}(\varepsilon_m^2)\,, \label{eq:evolutioneq2}
\end{align}
where $G_i$ corresponds to radiation reaction to the orbital elements of the binary, which drives the inspiral.

Note that the evolution equations \eqref{eq:evolutioneq1} and \eqref{eq:evolutioneq2} use the ``internal'' coordinate time  $t$ as the evolution parameter which is trivially redshifted by the ring with respect to the time $T$  of observers at infinity. It is also important to note that equations \eqref{eq:evolutioneq1} and \eqref{eq:evolutioneq2} represent the action-angle form of the evolution equations of an exactly integrable system at zeroth order in $\varepsilon_m$. Examples of such integrable systems considered in the EMRI scenario include bound geodesics in the Schwarzschild or the Kerr spacetime. Using canonical perturbation theory, however, we can approximate a nearly-integrable system by an integrable one, which is exactly what we did in the previous section. 

The global inspiral solution to the equations \eqref{eq:evolutioneq1} and \eqref{eq:evolutioneq2} can be naturally expanded with respect  $\varepsilon_m$ using a two-timescale analysis. The first timescale is the orbital timescale of the geodesic motion $\sim 1/\Omega$. Since this involves the evolution of the angles $\psi_i$ these are then called ``fast'' variables. On the other hand we have the inspiral timescale which deals with the decays of the actions on the much longer time-scale $\varepsilon_M J/G \sim \varepsilon_M/\Omega$. The actions are thus classified as ``slow'' variables. The standard procedure is then to separate these two timescales by averaging the functions $G_i$ over the fast variables, that is the $n$-dimensional invariant tori $\mathbb{T}^n$ parametrized by the angles
 \begin{align}\label{eq:averaging}
g_i(\mathbf{J})=\langle G_i(\mathbf{\psi},\mathbf{J})\rangle=\displaystyle\int_{\mathbb{T}^n}  G_i(\mathbf{\psi},\mathbf{J})\mathrm{d}^n\psi.
\end{align}

We can then first solve the equations for the  actions
 \begin{align}
\frac{\mathrm{d}J_i(t)}{\mathrm{d}t}=\varepsilon_m g_i(\mathbf{J}(t)),
\end{align}
while the angles can be obtain simply by integrating the fundamental frequencies whose evolution is given by the actions $J_i(t)$
\begin{align}\label{eq:angles}
\psi_i(\tilde{t}) =\displaystyle\int_{0}^{\tilde{t}} \Omega_i^{(t)} (\mathbf{J}({t})) \mathrm{d}t.
\end{align}

Let us now discuss the particular method we employed to compute the gravitational wave fluxes (i. e. the functions $g_i$). For this purpose we employ the quadrupole formalism which is the lowest order expansion in the post-Newtonian theory. The limitations of this method in the context of strong-field inspirals are obvious, but this method is sufficient for a qualitative analysis, and computing the fluxes with more sophisticated approximations is beyond the scope of our work. Hence, the flux formulas for the energy and the components of the angular momentum read
\begin{align}\label{eq:Quadflux}
\frac{\mathrm{d}E}{\mathrm{d}t}& =-\frac{1}{5} \sum_{i,j=1}^3 \bigg
\langle (\dddot I_{ij})^2 \bigg\rangle, \nonumber \\ \frac{\mathrm{d}L_i}{\mathrm{d}t} &=-\frac{2}{5} \sum_{j,k,l=1}^3 \epsilon_{ijk} \bigg\langle (\dddot I_{jl}\ddot I_{kl}) \bigg\rangle,
\end{align}
where the  traceless quadrupole moment of our particle has the form
\begin{align}\label{eq:radquadrup }
I^{ij}(t)=m \bigg( x^i(t) x^j(t)-\frac{1}{3}\delta^{ij} x^k(t) x_k(t)\bigg).
\end{align}
The functions $x^i(t)$ represent the orbital motion in Cartesian coordinates. Since we have so far used spherical-like coordinates, it is necessary to transform into the coordinates $x^i$, for that purpose we used the standard (flat-space) relations between spherical and Cartesian coordinates.

The time derivatives present in Eq.~\eqref{eq:Quadflux} are computed in accordance with the assumptions of the adiabatic approximation, which means that we neglect the change in the slow variables
\begin{align}\label{eq:fast derivative}
\dot x^i (\mathbf{\psi}(t),\mathbf{J}(t))=\frac{\partial x^i}{\partial \psi_j}\frac{\mathrm{d}\psi_j}{\mathrm{d}t}+\frac{\partial x^i}{\partial J_j}\frac{\mathrm{d}J_j}{\mathrm{d}t}\approx   \Omega_j^{(t)}\frac{\partial x^i}{\partial \psi_j}.
\end{align}
Note that the frequencies in the above expression are with respect to the coordinate time $t$ while our fundamental frequencies are related to the Mino time $\lambda$. This is not a problem since we have
\begin{align}
\Omega_j^{(t)} = \Omega_j \frac{\mathrm{d\lambda}}{\mathrm{d}t}.
\end{align}
In fact, we can exchange $t\rightarrow\lambda$ in higher time derivatives in the adiabatic approximation, since  the differentiation of $\frac{\mathrm{d\lambda}}{\mathrm{d}t}$ with respect to time would involve  terms proportional to the time derivatives of the actions which can be neglected as in \eqref{eq:fast derivative}. We can thus write
\begin{align}\label{timederivative}
\frac{\mathrm{d^n}}{\mathrm{d}t^n}\approx\bigg(\frac{\mathrm{d\lambda}}{\mathrm{d}t}\bigg)^n \frac{\mathrm{d^n}}{\mathrm{d}\lambda^n},\hspace{35pt} \frac{\mathrm{d}t}{\mathrm{d}\lambda}=\frac{\partial H_N}{\partial \delta E} (\mathbf{J}).
\end{align}

We would like to remind the reader here that all quantities depending on $\mathbf{J}$ also depend on $E$ (or $\delta E$), but the energy and the three actions are not independent, since we have the normalization condition~\eqref{eq:Normalization}, which is why the energy dependence is often omitted. Ideally the coordinate functions $x^i(t)$ can be expressed as a Fourier-like expansions, the same can then be said about expression for the fluxes. The averaging is then equivalent to eliminating all the oscillating terms 
\begin{align}\label{eq:analytaver}
G =\displaystyle\sum_{k} c_{\mathbf{k}}(\mathbf{J}) e^{i(\mathbf{k}\cdot \mathbf{\psi}(t))} \enspace  \Rightarrow\enspace \big\langle G \big\rangle =c_{\mathbf{0}}(\mathbf{J}),\hspace{15pt} \mathbf{k} \in \mathbb{Z}^3.
\end{align}

In practice, however, this fully  analytical approach is not feasible because of the number of terms present in Eq.~\eqref{eq:Quadflux}, this is especially true for the non-equatorial orbits. It is easier to numerically integrate the function $G$.  Instead of using a multidimensional integral like in Eq.~\eqref{eq:averaging}, we can integrate over an orbit that densely covers the invariant torus determined by the actions. This implies integrating over a sufficiently long time $\Lambda$
\begin{align}\label{eq:numeraver}
\langle G(\mathbf{\psi}(\lambda),\mathbf{J}) \rangle=\frac{1}{\Lambda} \int_0^{\Lambda} G(\mathbf{\psi}(\lambda),\mathbf{J})\mathrm{d}\lambda .
\end{align}

We now have to find the evolution equations for the three independent integrals of motion. It is straightforward to use the energy and the $z$-component of the angular momentum ($J_\phi$) since we have explicit formulas \eqref{eq:Quadflux} for them. The third integral will be the action $J_\theta$, which can be written as
\begin{align}\label{thetaaction}
 J_\theta=J^{(0)}_\theta -\delta J_\theta, \hspace{15pt}  \delta J_\theta= \lbrace J_\theta , \chi_1 \rbrace ,
\end{align}
where the action $J^{(0)}_\theta$ is the original one derived in Eq.~\eqref{eq:AAIsol5}, i.e. the one before applying the Lie operator with the generating function $\chi_1$ (see Eq.~\eqref{eq:gf_exp}). Knowing the quadrupole fluxes for the angular momentum components  we can compute its time derivative as
\begin{align}
 \frac{\mathrm{d}J^{(0)}_\theta}{\mathrm{d}t}=\frac{\vec{L}\cdot\frac{\mathrm{d}\vec{L}}{\mathrm{d}t}}{L}-\frac{\mathrm{d}J_\phi}{\mathrm{d}t}.
\end{align}

The components of the angular momentum can be expressed in terms of our original phase-space coordinates as 
\begin{align}\label{angularcompon}
L_x= -\sin \left( \phi \right) p_\theta -\cos \left( \phi \right) \cot
 \left( \theta \right) {\it J_\phi}, \\
 L_y= \cos \left( \phi \right) p_\theta -\sin \left( \phi \right) \cot
 \left( \theta \right) {\it J_\phi}.
\end{align}
Since the function $\delta J_\theta$ contains only oscillating terms it does not survive the averaging
\begin{align}
\bigg\langle \frac{\mathrm{d}J_\theta}{\mathrm{d}t} \bigg\rangle=\bigg\langle \frac{\mathrm{d}J^{(0)}_\theta}{\mathrm{d}t}\bigg\rangle, \hspace{15pt} \bigg\langle \frac{\mathrm{d}\delta J_\theta}{\mathrm{d}t} \bigg\rangle\rightarrow 0 \hspace{10pt} \text{as}  \hspace{10pt} \Lambda\rightarrow \infty.
\end{align}
We, thus, arrive at the complete system of evolution equations for three independent integrals of motion
These three equations can then be solved numerically, The radial action $J_r$ can be computed at each time step from the normalization condition~\eqref{eq:Normalization}, while the evolution of the angles is given by the integrals of their respective fundamental frequencies (expression~\eqref{eq:angles}). 

The computation of the fluxes is detailed in the second Maple notebook \cite{SupMat2}

\section{Results} \label{sec:results}

In this section we present the adiabatic evolution in the  general nonequatorial case. Let us again stress that our results provide essentially a qualitative analysis due to the approximative methods we have employed. This includes the particular values of various parameters we have used in this section, some of which are not relevant for realistic EMRIs. For instance, the mass ratio we use in this section is $\varepsilon_m=10^{-3}$, but it does not practically matter in our approach, since the equations which govern the adiabatic evolution of actions do not depend explicitly on the time; any other value of $\varepsilon_m$ would  just rescale the time variable in our solution. Another important parameter is the external quadrupole $Q$ representing the ring, its value was intentionally chosen to be large ($Q=10^{-6}$) in the following so that its effect is prominent in the figures. Lastly the ring radius is set to $r_{\rm r}=50 M$. 

\begin{figure*}[ht!] \centering
\begin{subfigure}[b]{0.475\linewidth}
	\includegraphics[width=\linewidth]{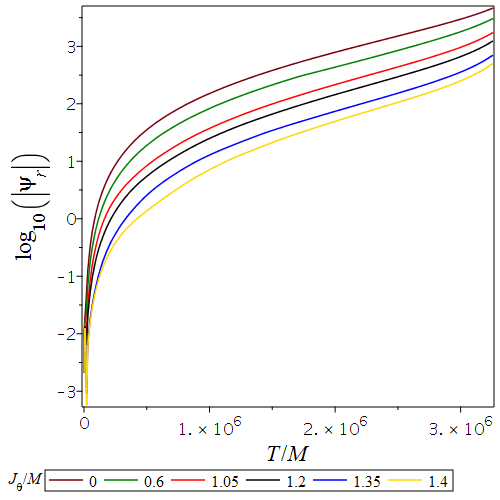}
	\end{subfigure}
\begin{subfigure}[b]{0.475\linewidth}
	\includegraphics[width=\linewidth]{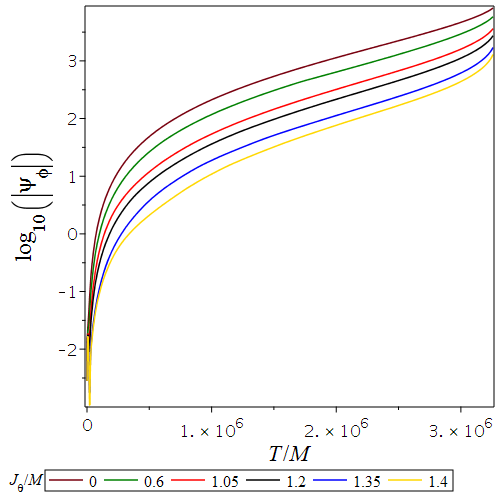}
	\end{subfigure}
\caption{ The left panel shows logarithmic plot of the radial phase shift $\delta\psi_r(T)$  (left panel) and the right panel shows logarithmic plot of the azimuthal one $\delta\psi_\phi(T)$ for different initial value values of $J_\theta$. For these plots we have used $Q=10^{-6}M^{-2}$ and $r_{\rm r}=50 M$ while the initial frequencies are matched to the Schwarzschild ones with  $J_r(0)=0.1M, L(0)=5M$. In all cases the phase shifts remain negative during the evolution.}
\label{fig:shifts}
\end{figure*}

\subsection{Phase shifts}

First we investigate the effect of the ring perturbation on the orbital phases, i.e. the angle coordinates.
\begin{align}\label{shifts}
\delta\psi_i(t)= \psi_i(t)-\psi_i(t)\vert_{Q=0}, \,
\end{align}
where the evolution of $\psi_i(t)$ is given by Eq.~\eqref{eq:angles}. As the fundamental frequencies are in principle observable, it is natural to parametrize our orbits by them instead of the actions. Thus, we start from the same initial frequencies in both the perturbed and unperturbed cases so that not only $\delta\psi_i(0)=0$, but also  $\delta \dot \psi_i(0)=0$. This matching of the frequencies was used  in \cite{freqmatch} in the case of a spinning particle in Kerr spacetime. Unlike in their case however our reference spacetime is Schwarzschild  where  $\Omega_\theta^{Schw} = \Omega_\phi^{Schw} $ which is a consequence of spherical symmetry while for $Q \neq 0$ we have $\Omega_\theta  \neq \Omega_\phi $. \par
Despite the inability to match all the frequencies we can still choose two of them (in our case  $\Omega_r$ and $\Omega_\phi$) and match them to their  Schwarzschild counterparts for a fixed value of $Q$. In addition we can find the matches for different values of $J_\theta$, which effectively means different initial inclinations.

When evolving the angles (or other quantities) one should use the proper time $T$ of the asymptotic observer as an evolution parameter instead of $t$. This involves including the redshift factor $z_{in}$  given by Eq.~\eqref{redshift}, which was absorbed into the coordinates $t$ and $r$. This monopole term of the expansion (see Eq.~\eqref{potenexpansion}) is necessary to include what is dynamically dominating; however, it is the non-constant quadrupole term which breaks the spherical symmetry. For this reason, we compute the phase shift only for the quadrupole perturbation which means using the definition \eqref{shifts} but with the time $T$. The  phase shifts for the two matched frequencies are then plotted in the Fig.~\ref{fig:shifts}.  It is interesting to see that $\vert\delta\psi_i\vert$ is smaller for larger values of $J_\theta$, keep in mind that $\psi_i(t)\vert_{Q=0}$  remains  unchanged as the evolution in the Schwarzschild spacetime does not depend on the initial inclination.

\subsection{Inclination and eccentricity}

\begin{figure*}[ht!] \centering
\begin{subfigure}[b]{0.35\linewidth}
	\includegraphics[width=\linewidth]{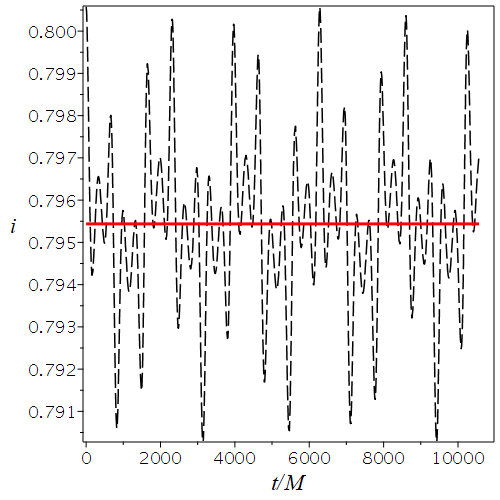}
	\end{subfigure}
\begin{subfigure}[b]{0.35\linewidth}
	\includegraphics[width=\linewidth]{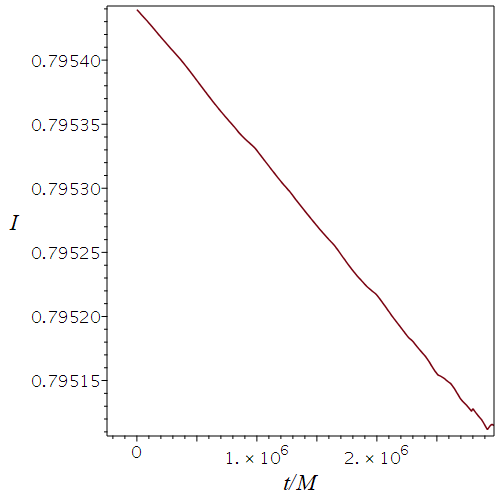}
	\end{subfigure}
\caption{The left panel shows the oscillation of the orbital inclination (dashed black)  with respect to the averaged constant inclination (red continuous curve) for a geodesic orbit. The right panel shows the evolution of the averaged inclination $I$ during the inspiral. For these plots we have used $Q=10^{-6} M^{-2}, J_r(0)=0.11M,  J_\theta(0)=1.5M, J_\phi(0)=3.5M$.  }
\label{fig:inclination}
\end{figure*}

When considering non-equatorial motion in a non-spherically symmetric spacetime one can study the behavior of  orbital inclination $i$. This quantity is defined as an angle between the current orbital plane and the equatorial plane. In terms of our action variables it can be expressed as
\begin{align}\label{eq:inclination}
i = \arccos \bigg(\frac{J_\phi}{J_\phi+J^{(0)}_\theta}\bigg).
\end{align} 
When the quadrupole perturbation is present the inclination of a geodesic orbit does not remain constant, but it oscillates. The oscillations is caused by the $\delta J_\theta$ term contained in $J^{(0)}_\theta$ (see Eq.~\eqref{thetaaction}).

In order to show the effect of the adiabatic evolution on the inclination, it is useful to separate this geodesic evolution by defining the averaged inclination
\begin{align}\label{eq:averaged inclination}
 I=\langle i\rangle = \arccos \bigg(\frac{J_\phi}{J_\phi+J_\theta}\bigg).
\end{align}
This inclination is constant in the geodesic case, since it depends only on the integrals of motion. During an inspiral, however, $I$ shall evolve on the inspiral timescale and it is interesting to compare  the geodesic oscillation of $i$ to the EMRI evolution of $I$.

\begin{figure}[h!] \centering
	\includegraphics[width=0.9\linewidth]{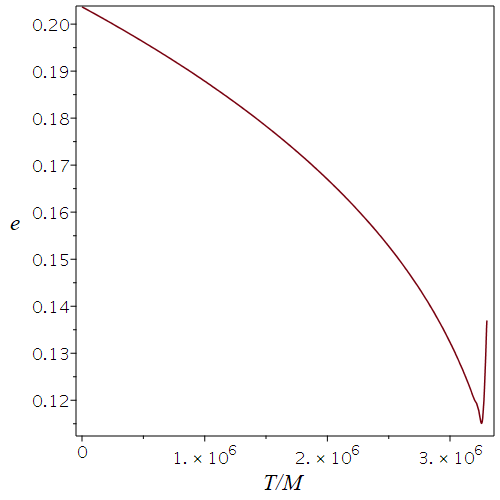}
\caption{The adiabatic evolution of eccentricity as a function of time $T$ in the case of an equatorial inspiral ($Q=10^{-6} M^{-2}, J_r(0)=0.098M,  J_\theta(0)=0, J_\phi(0)=5M$.).}
\label{fig:eccentricity}
\end{figure}

Fig.~\ref{fig:inclination} shows that the geodesic oscillation of inclination for a particular choice of initial conditions is two orders of magnitude larger than the drift of $I$ caused by radiation reaction. This difference of course depends on the strength of the perturbation ($Q$) which in our example is quite large, however, we must also keep in mind that the influence of the ring on the dynamics decreases as we approach the black hole.  This is caused by the smaller value of the total angular momentum close to the ISCO, where the amplitude of the geodesic oscillations of $i$ is comparable to the total change of $I$ during the EMRI.  The fact that the function $I$ is decreasing is expected, as the dissipation of the constants of motion should, in principle, lead to the equatorial plane value $I=0$. On the other hand, for some initial conditions we have seen an increase of $I$ as the inspiral reaches ISCO, we should speculate this effect to be possibly of numerical origin. This growth is also present in the case of  eccentricity, which can be defined as  
\begin{align}\label{eq:ecc}
 e=\frac{r_1-r_2}{r_1+r_2}
\end{align}
where $r_1$ is the maximum value of $r(t)$ for a given geodesic while $r_2$ is the corresponding minimum. An evolution of the eccentricity during the inspiral  can be seen in Fig.~\ref{fig:eccentricity}. The growth of eccentricity close to the ISCO was  also found in other works (see, e.g., \cite{eccgrowth}),  but it is questionable whether it has a physical significance or it is just a coordinate effect.

\subsection{Waveforms}

\begin{figure*}[ht!] \centering
\begin{subfigure}[b]{0.45\linewidth}
	\includegraphics[width=\linewidth]{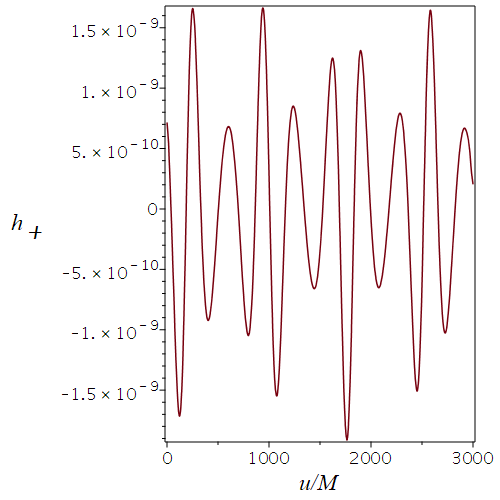}
	\end{subfigure}
\begin{subfigure}[b]{0.45\linewidth}
	\includegraphics[width=\linewidth]{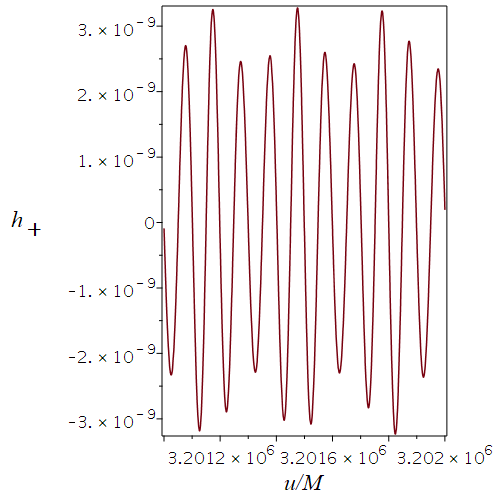}
	\end{subfigure}
\caption{The gravitational wave strain ( $h^{\lbrace+\rbrace}$) of a single non-equatorial EMRI ($Q=10^{-6} M^{-2}, J_r(0)=0.11M,  J_\theta(0)=1.5M, J_\phi(0)=3.5M$) at two different instants: $u=0$ (left panel) and $u=3.2\cdot 10^6 M$ (right panel)}
\label{fig:waveform}
\end{figure*}

Finally let us conclude our results with some plots of  gravitational waveforms. In our radiation-quadrupole formalism the components of the metric perturbation can be written in the TT gauge  as
\begin{align}\label{eq:hssol0}
h_{ij}^{TT}=\frac{2}{R} \ddot I_{ij}^{TT}(u).
\end{align}
where $u=T-R$ is the retarded time and $I_{ij}^{TT}$  can be obtained from $I_{ij}$ using projectors $ P_{ij}$ as
\begin{align}\label{eq:hssol1}
I_{ij}^{TT}=P_{i}^{k} I_{kl} P^{l}_{j}-\frac{1}{2} P_{ij} P^{kl} I_{kl}, \quad P_{ij}=\delta_{ij}-n_i n_j.
\end{align}
where $n_i$ is a unit vector pointing from  source to the observer. Naturally, as in the case of the calculations of the fluxes, $I_{kl}$ depends on the coordinates of our particle. Their adiabatic evolution is determined by the evolution of actions and angles 
\begin{align}
x^i(T)= x^i(\psi(T),\mathbf{J}(T)). 
\end{align}
In following we  decompose  $h_{ij}^{TT}$  as it was done in~\cite{Moore}. For that we need  to define two additional vectors $\vec{p}$ and $\vec{q}$ which together with $\vec{n}$ form an orthonormal basis in the 3D Euclidean space.
\begin{align*}
    \vec{p} = \frac{\vec{n} \times \vec{L}}{\vert \vec{n} \times \vec{L} \vert},\hspace{35pt} \vec{q}=\vec{p} \times \vec{n}.
\end{align*}
The components of the angular momentum $\vec{L}$ can be computed from our action-angle variables as was the case in  Eq.~\eqref{angularcompon}. The two independent polarizations have the form
\begin{equation}\label{eq:hssol2}
h^{\lbrace+,\times\rbrace}=\frac{1}{2} H^{\lbrace+,\times\rbrace}_{ij} h^{ij}_{TT} ,
\end{equation}
where $H^{\lbrace+,\times\rbrace}_{ij}$ are defined as
\begin{equation}\label{eq:hssol3}
H^{+}_{ij}=p_i p_j-q_i q_j,\hspace{35pt}  H^{\times}_{ij}=p_i q_j-q_i p_j. 
\end{equation}

With all the ingredients in place we can plot some waveforms. One such an example is depicted in Fig.~ \ref{fig:waveform}, where we can see the component $h^{\lbrace+\rbrace}$ at the beginning of the evolution around $r\approx 21M$  and at a later time when $r\approx 9M$. Despite not having decomposed the signal into modes it is evident from  the figure that the amplitudes and frequencies grow during the inspiral as expected.

\begin{figure}[h!] \centering
	\includegraphics[width=0.95\linewidth]{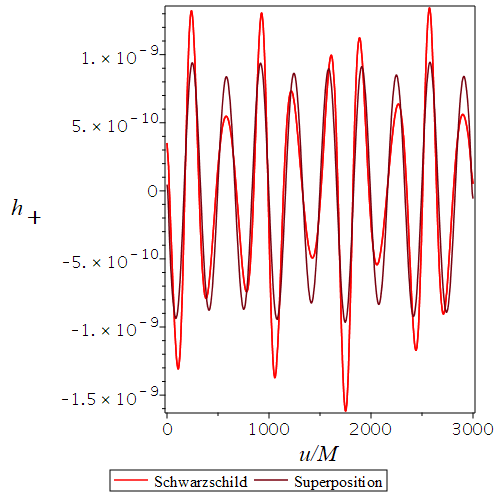}
\caption{A comparison of the unperturbed (red) and perturbed waveform for two matched frequencies in the equatorial plane ($Q=10^{-6} M^{-2}, J_r(0)=0.002M,  J_\theta(0)=0 , J_\phi(0)=5M$.) }
\label{fig:wavecomp}
\end{figure}

In Fig.~\ref{fig:wavecomp} we compare two waveforms to see the effect of the quadrupole term in our Hamiltonian. Although the initial radial and azimuthal frequencies are matched as above the phase shifts tend to grow rather quickly for the large value of $Q$ we had chosen. In addition to that we can see a great difference in the amplitudes as well.

Throughout the Sec.~\ref{sec:results}  we included the value $Q$ for each figure. It is, however, the quantity $Q L^4/M^2$ which characterizes the strength of the quadrupole perturbation (as was mentioned in Sec.~\ref{sec:validity}). The reason we chose $Q$ over $Q L^4/M^2$ is because the latter is not a constant during  the inspiral as the total angular momentum $L$ is a decreasing function of time. Namely, the effect of perturbation are lower as we get farther from the ring. On the other hand, it is important to point out that due to the expansion scheme we used the effects of the perturbation will again grow close to the Schwarzschild's ISCO, where the approximation breaks down as was demonstrated in Sec.~\ref{sec:validity}  for the precession  rates. Nevertheless, the maximum strength of the perturbation is at the beginning of the adiabatic evolution, where we had $Q L^2(0)/M^2 = 2.5 \cdot 10^{-5}$.

\section{Summary and discussion} \label{sec:Concl}

This work showcased the advantages of using the Lie series approach to tackle the EMRI problem. For this purpose, we used a fairly complex background system, in which the primary black hole is surrounded by a matter distribution. In particular, we truncated a gravitating ring-like source up to its leading quadrupole term to introduce a quite generic tidal field around the primary Schwarzschild black hole. We wrote the Hamiltonian system providing the geodesic motion in the above background and noticed that it can be split into a part giving the motion in the Schwarzschild background, which correspond to an integrable system, and a perturbative part expressing the perturbation due to the quadrupole term.  

By using the Mino time, we further split the Schwarzschild part of the Hamiltonian into a radial and angular part. After relatively simple manipulation we showed that the angular part can be written in action-angle variables, while for the radial part we perturbed it around a circular orbit as a harmonic oscillator and used a standard canonical transformation to get it into action-angle variables as well. To expand our scheme further from the circular orbit we applied two canonical transformations using the Lie series approach on the Schwarzschild part of the Hamiltonian. The same series of transformations were also applied on the perturbative part of the Hamiltonian leading to a Hamiltonian system in action-angle variables valid up to the separatrix. We tested the obtained Hamiltonian system and found that as far as we stay away from the $1:1$, $1:2$ and $2:3$ resonances between the radial and polar frequencies and ISCO the system is behaving sufficiently well.  

After establishing the conservative part of our approximation to an EMRI, we addressed the dissipative part. To introduce dissipation into the system we used fluxes computed by the quadrupole formula. By averaging out the oscillating terms of the fluxes, we were able to provide the equations for the adiabatic radiative decay of the actions and evolve the inspirals. Since we derived the characteristic frequencies of the system, we were able to easily obtain the orbital phase shifts caused by the matter distribution. Moreover, we were able to compute the eccentricity and inclination changes as the inspiral evolves and the respective waveforms.

In the future, we would like to improve this work in a number of ways. First, it is necessary to also treat the case of a perturbed inspiral into a generic spinning black hole, that is the Kerr space-time. Second, our formalism breaks down near resonances, so we would like to implement a variant of the formalism sketched in Ref. \cite{Lukes-Gerakopoulos2020} to evolve the inspiral faithfully through the resonances. Third, we have restricted to the case of a perfectly axially symmetric stationary cloud of matter. However, in astrophysically realistic scenarios the external matter sources are only approximately so. In particular, when the external matter consists of a halo of orbiting objects such as stars, the largest deviations from stationarity and axisymmetry come from those objects that have either outstanding masses or are very close to the center \cite{Bonga19}. Even though the question of resonances caused by such perturbations was already treated by Refs. \cite{Bonga19,gupta2021importance}, we wish to systematically address the symmetry breaking within our formalism in the future.

The last, but perhaps most salient point we would like to improve upon in the future is the question of the gravitational-wave fluxes of energy and angular momentum. Rather obviously, it is necessary to include a strong-field flux computation using the Teukolsky equation or a similar method. However, the additional issue is that the tidal quadrupole perturbation also causes a perturbation to the Teukolsky equation. This perturbation then adds a $Q$-proportional contribution to the flux, which implies a comparable contribution to the inspiral phasing as the perturbation to the geodesics we have treated here. However, the perturbation makes the equation non-separable and will require a delicate analysis. 

\section{Acknowledgements}

LP and GLG have been supported by the fellowship Lumina Quaeruntur No. LQ100032102 of the Czech Academy of Sciences. VW was supported by by European Union’s Horizon 2020 research and innovation programme under grant agreement No 894881. LP acknowledges support by the project ”Grant schemes at CU” (reg.no.CZ.02.2.69/0.0/0.0/19 073/0016935).

\bibliographystyle{unsrt}
\bibliography{refs}

\appendix

\section{Derivation of perturbed black-hole field} \label{app:ring}

To derive the tidally perturbed black hole field, we use the formulas for black hole fields surrounded by light ring-like sources at finite distances as recently presented by {\v C}{\'i}zek \& Semer{\'a}k \cite{Cizek:2017wzr} (see also the seminal work of Will \cite{Will:1974zz}). They start from metrics of the form
\begin{align}
\begin{split}
    & \d s^2  = -e^{2 \nu} \d T^2 + R^2\left(1 - \frac{M^2}{4 R^2}\right)^2 e^{-2\nu} (\d \phi - \omega \d T)^2 
    \\ & \phantom{\d s^2  =}+ e^{2 \zeta - 2 \nu} (\d R^2 + R^2 \d \theta^2)\,,
\end{split}
\end{align}
where $T,\phi,R,\theta$ are coordinates of the Carter-Thorne-Bardeen type, and $\nu,\omega,\zeta$ are unknown metric functions. The zeroth-order solution (isolated static black hole) is presented in this case by the metric functions
\begin{align}
    \nu_0 = \ln \left( \frac{2 R - M}{2 R + M} \right), \,
    \omega_0 = 0\,,\,
    \zeta_0 = \ln \left(1 - \frac{ M^2}{4 R^2}\right).
\end{align}
It can then be easily seen that $R$ is the isotropic radius at zeroth order. The linear perturbations $\nu = \nu_0 + \delta\nu,\, \omega =\omega_0 + \delta \omega$ by a rotating ring are then obtained by using Green's functions $\mathcal{G}^\nu$ and $\mathcal{G}^\omega$ in equations (66) and (75) of {\v C}{\'i}zek \& Semer{\'a}k \citep{Cizek:2017wzr}. Specifically, for a ring of Komar mass $\mathcal{M}_{\rm r}$ and angular momentum $\mathcal{J}_{\rm r}$ we obtain 
\begin{align}
    & \delta \nu = -\frac{2 \mathcal{M}_{\rm r}}{M} \mathcal{G}^\nu (x(R),\theta,x(R_{\rm r}),\pi/2)\,, \\
    & \delta \omega =  -\frac{8 \mathcal{J}_{\rm r}}{M^3} \mathcal{G}^\omega(x(R),\theta,x(R_{\rm r}),\pi/2)\,,\\
    & x(R) = \frac{R}{M}\left(1 + \frac{M^2}{4 R^2}\right) .
\end{align}
where $x(R)$ is the auxiliary dimensionless radius used by {\v C}{\'i}zek \& Semer{\'a}k. The perturbation to $\zeta = \zeta_0 + \delta \zeta$ is then obtained by a particular line integral involving the gradient of $\nu$. We expand the Green's function in the limit $R_{\rm r} \gg R \sim M$ to obtain
\begin{align}
\begin{split}\label{potenexpansion}
    & \mathcal{G}^\nu = \frac{M}{2 R_{\rm r}} 
    \\& \phantom{\mathcal{G}^\nu =} + \frac{M\left[(M^2 + 4 R^2)^2 - (3 M^4 + 8M^2 R^2 + 48R^2) \cos^2\!\theta\right]}{128 R^2 R_{\rm r}^3} 
    \\& \phantom{\mathcal{G}^\nu =}+ \mathcal{O}(R_{\rm r}^{-4})\,, 
    \end{split}\\
    & \mathcal{G}^\omega = -\frac{M^3}{4 R_{\rm r}^3} + \frac{3 M^4}{4 R_{\rm r}^4} + \mathcal{O}(R_{\rm r}^{-5})\,.
\end{align}
Note that we consider the $\mathcal{G}^{\omega}$ expansion to order $R_{\rm r}^{-4}$ since it enters the metric multiplied by $\mathcal{J}_{\rm r} \sim R_{\rm r}^{1/2}$. Finally, we obtain for $\delta \zeta$ 
\begin{align}
    \delta \zeta =  -{\frac {\mathcal{M}_{\rm r} M  \sin^2\!  \theta \left( {M}
^{2}+4\,{R}^{2} \right) }{4\,R \, R_{\rm r}^3}} + \mathcal{O}(R_{\rm r}^{-4})\,.
\end{align}

Now the transformation to the local Schwarzschild-like coordinates in which the metric attains the form \eqref{eq:metric} is given by
\begin{align}
    & r = R \left(1 + \frac{M}{2R}\right)^2 (1+z_{\rm in}), \,\th = \theta\,,\\
    & t - t_0 = T(1 - z_{\rm in})\,, \ph- \ph_0 = \phi - \Omega_{\rm in} T\,,  
\end{align}
where $t_0,\ph_0$ are integration constants and the redshift and angular-velocity factors are
\begin{align}
    & z_{\rm in} = \frac{ \mathcal{M}_{\rm r}}{R_{\rm r}} = \frac{ \mathcal{M}_{\rm r}}{r_{\rm r}} \left(1 + \frac{M}{r_{\rm r}}\right) + \mathcal{O}(r_{\rm r}^{-3}) \,,\\
    & \Omega_{\rm in} = \frac{2 \mathcal{J}_{\rm r}}{R_{\rm r}^3}\left(1 - \frac{3 M}{R_{\rm r}}\right) =\frac{2 \mathcal{J}_{\rm r}}{r_{\rm r}^3} + \mathcal{O}(r_{\rm r}^{-5})\,.
\end{align}
%

\section{Schwarzschild in action-angle coordinates}\label{app:Schw}

In order to find the action-angle form of the Schwarzschild Hamiltonian we first have to separate its radial and angular parts which can be done by replacing the proper time as an evolution parameter using  $\mathrm{d}\tau=r^2\mathrm{d}\lambda$. One can easily make sure that the Hamiltonian $H_{(\lambda)}=\frac{1}{2}r^2 (2~H+1)=\frac{1}{2}r^2 (g^{\mu\nu}p_\mu p_\nu+1)$ is the generator of evolution in $\lambda$
\begin{align*}
 \frac{d x^\mu}{d \lambda} &= \frac{\partial H_{(\lambda)}}{\partial p_\mu}=r^2\frac{\partial H}{\partial p_\mu}=r^2\frac{d x^\mu}{d \tau}\, , \\
 \frac{d p_\theta}{d \lambda} &= -\frac{\partial H_{(\lambda)}}{\partial \theta}=-r^2\frac{\partial H}{\partial \theta}=r^2\frac{d p_\theta}{d \tau}\, ,\\
  \frac{d p_r}{d \lambda} &= -\frac{\partial H_{(\lambda)}}{\partial r}=-r^2\frac{\partial H}{\partial r}-r(2~H+1)=r^2\frac{d p_r}{d \tau}\, .
\end{align*}

The angular part of the Hamiltonian takes a simple form \eqref{eq:angularH} with actions $J_\phi=p_\phi$  and  $J_\theta=L-J_\phi$ (derived using the integral \eqref{eq:AAIsol5} ) the angles can be found using generating function of second kind which is a solution to the corresponding Hamilton-Jacobi equation. We can take advantage of the separability of the  Hamilton-Jacobi equation and write the generating function as 
\begin{align}
 S=S_\theta(\theta,J_\theta,J_\phi)+S_\phi(\phi,J_\phi)= \int p_\theta(\theta,J_\theta,J_\phi) \mathrm d \theta+ \phi J_\phi.
\end{align}
From here it is straightforward to get the angles conjugated to actions $J_\theta$ and  $J_\phi$
\begin{align}
\psi_\theta  =  \frac{\partial S}{\partial J_\theta},\hspace{35pt} \psi_\phi  =  \frac{\partial S}{\partial J_\phi}.
\end{align}
These expressions  can then be inverted to express the old coordinates in terms of the new. For the coordinates $\theta$ and $p_\theta$ we have
\begin{align}\label{eq:thvar2aa}
\theta &=\pi -\arccos \left( \sqrt {1-{\frac {{J_\phi}^{2}}{ \left( J_\phi+{\it J_\theta
} \right)^2}}}\sin \left( \psi_\theta \right)  \right), \\
 p_\theta &=\cos \left( \psi_\theta \right) \sqrt {{\frac {{\it J_\theta}\, \left( {{\it J_\theta^3}}+4\,{{\it J_\theta^2}}J_\phi+5\,{\it J_\theta}\,{J_\phi^2}+2\,{J_\phi^3} \right) }{{{
\it J_\theta^2}}  \cos^2 \left(\psi_\theta   \right) +2\,{\it J_\theta
}\,J_\phi \cos^2 \left( \psi_\theta \right)  +{J_\phi^2}}}}, \nonumber
\end{align}
while the coordinate $\phi$ can be written as 
\begin{align}\label{eq:phi}
\phi &=\psi_\phi-\psi_\theta \nonumber \\
     &+{\frac {1}{2}\arctan \left( {\frac {J_\phi \left( J_\phi+{\it J_\theta} \right) \sin
 \left( 2\,\psi_\theta \right) }{  \cos^2 \left( \psi_\theta \right)   
 \left( {{\it J^2_\theta}}+2\,{\it J_\theta}\,J_\phi+2\,{J^2_\phi} \right) -{J^2_\phi}}
} \right) }.
\end{align}
Of course one has to keep in mind that in this expression it is necessary to add factor $\pi/2$ each time the denominator inside $\arctan$ is zero so that the transformation is continuous. It is also worth noting that in the equatorial plane ($J_\theta=0$) the equation \eqref{eq:phi} is reduced to $\phi=\psi_\phi$.

The following step is to write  the radial Hamiltonian \eqref{eq:Hrad} as a Taylor expansion  from a stable circular orbit. The location of a circular orbit ($r=r_c$)  is related to the total angular momentum $L=J_\theta+J_\phi$  as
\begin{align}
r_c={\frac {L}{2\,M} \left( L+\sqrt 
{  L^{2}-12\,{M}^{2}} \right) }.
\end{align}

Our expansion parameter is $\varepsilon$. The order of $\varepsilon$ is for the relevant quantities given by \eqref{eq:expansion}. In particular when expanding the energy one obtains
 \begin{align}\label{eq:Energyexp} 
E= E_c+\delta E &=E_c+\frac{\partial E }{\partial r}\Big\vert_{\substack{r=r_\mathrm{c} \\ p_r=0}}(r-r_\mathrm{c})+\frac{\partial E }{\partial p_r}\Big\vert_{\substack{r=r_\mathrm{c} \\ p_r=0}} p_r+\mathcal{O}(\varepsilon^{2}) \nonumber \\
 &=E_c+\mathcal{O}(\varepsilon^{2}),
\end{align}
where the partial derivatives vanish since our stable circular orbit has minimal energy, thus, we get $\delta E=\mathcal{O}(\varepsilon^{2})$. The energy of the circular orbit is then
  \begin{align}
E_c=\frac {r_{{c}}-2\,M}{\sqrt{ r_c^{2}-3\,Mr_c}}.
\end{align}

We can now expand the radial  Hamiltonian, identify the harmonic oscillator terms and transform them into action-angle coordinates
\begin{align}\label{eq:AAIsol9}
r&={\it r_\mathrm{c}}+\sqrt {{\frac {2J_r{{\it r_\mathrm{c}}}^{2}}{\Omega_{rc}} \left( 1-
\,{\frac {2M}{{\it r_\mathrm{c}}}} \right) }}\sin \left( \psi_r \right) , \nonumber \\
{\it p_r}&=\sqrt {{\frac {2J_r\Omega_{rc}}{{{\it r_\mathrm{c}}}^{2}\left( 1-
\,{\frac {2M}{{\it r_\mathrm{c}}}} \right)}}} \cos \left( \psi_r \right) ,
\end{align}
where the frequency of the harmonic oscillator can be written as 
\begin{align}
\Omega_{rc}=\sqrt{{\frac {M \left( {\it r_c}-6\,M \right) {\it r_c}}{r_c-3\,M
}}}
.
\end{align}

After performing the transformation we arrive at the radial Hamiltonian in the form
\begin{align}\label{eq:Hradial}
  H_{\rm rad} &=\frac{1}{2}{\frac {M{{\it r_\mathrm{c}}}^{2}}{3\,M-{\it r_\mathrm{c}}}}-{\frac {{{\it r_\mathrm{c}}}^{3}
\delta E}{\sqrt {{\it r_\mathrm{c}}\, \left(r_\mathrm{c}- 3\,M\right) }}}+J_r\Omega_{rc} \nonumber \\
   &+R(\delta E,\psi_r,J_r).   
 \end{align}
We can now employ the canonical perturbation theory to get farther from the circular orbit and closer to the separatrix. In our case we applied the Lie operator twice 
\begin{align}
\exp( \pounds_{\omega_{2}})\exp( \pounds_{\omega_{1}})H_{{\rm Schw}(\lambda)}=H_{NS}(J_r,J_\theta)+\mathcal{O}(\varepsilon^{5}).
\end{align}
For instance the first generating function $\omega_{1}$ can be written as
\begin{widetext}
\begin{multline*}
\omega_1=\frac {4\,\sqrt {2}}{3\, \left( -r_{{c}}+2\,M \right) ^{4}{\Omega}^{2
}fr_{{c}}} \Bigg\lbrace  \left(  \left( -{\frac {r_{{c}}}{2}}+M \right) ^{4}
 \left( M-r_{{c}} \right) {\Omega}^{2}+{\frac {{r_{{c}}}^{4}{f}^{2}{M}
^{3}{E_{{c}}}^{2}}{4}} \right) J_{{r}}{{\rm e}^{-3\,i\psi_{{r}}}} +\\ +
 \left( 3\, \left( -\frac{1}{2}\,r_{{c}}+M \right) ^{4} \left( M-r_{{c}}
 \right) J_{{r}}{\Omega}^{2}-9\,{r_{{c}}}^{4}f \left( -\frac{1}{2}\,r_{{c}}+M
 \right) ^{2}\delta E\, \left( M-\frac{1}{3}\,r_{{c}} \right) E_{{c}}\Omega-{
\frac {9\,{r_{{c}}}^{4}{f}^{2}J_{{r}}{M}^{3}{E_{{c}}}^{2}}{4}}
 \right) {{\rm e}^{-i\psi_{{r}}}}+\\ + \left(  \left( -{\frac {r_{{c}}}{2}
}+M \right) ^{4} \left( M-r_{{c}} \right) {\Omega}^{2}+{\frac {{r_{{c}
}}^{4}{f}^{2}{M}^{3}{E_{{c}}}^{2}}{4}} \right) J_{{r}}{{\rm e}^{3\,i
\psi_{{r}}}}+\\ +3\, \left(  \left( -\frac{1}{2}\,r_{{c}}+M \right) ^{4} \left( M-
r_{{c}} \right) J_{{r}}{\Omega}^{2}-3\,{r_{{c}}}^{4}f \left( -\frac{1}{2}\,r_{
{c}}+M \right) ^{2}\delta E\, \left( M-\frac{1}{3}\,r_{{c}} \right) E_{{c}}
\Omega-\frac{3}{4}\,{r_{{c}}}^{4}{f}^{2}J_{{r}}{M}^{3}{E_{{c}}}^{2} \right) {
{\rm e}^{i\psi_{{r}}}} \Bigg\rbrace \sqrt {{\frac {J_{{r}}f}{\Omega}}},
\end{multline*}
\end{widetext}
where $f$ is  the factor  Schwarzschild factor
\begin{align}
f=1-\,{\frac {2M}{r_{{c}}}}.
\end{align}

The normal form of the Schwarzschild Hamiltonian reads up to the $\mathcal{O}(\varepsilon^5)$ terms 
\begin{widetext}
\begin{multline*}
H_{NS}={\frac {M{{\it r_c}}^{2}}{6\,M-2\,{\it r_c}}}-{\frac {{{\it r_c}}^{3}
\delta E}{\sqrt { \left| {\it r_c}\, \left( 3\,M-{\it r_c} \right) 
 \right| }}}+J_r\Omega+\frac{1}{2}\, \left( {\it J_\theta}+L_z \right) ^{2}+\\+ 
\frac{1}{4{\it r_c}\,{\Omega}^{4} \left( -{\it r_c}+2
\,M \right) ^{5}} \,\bigg\lbrace48\, \left( {J_r}^{2}{M}^{2}-\frac{4}{3}\,M{\it r_c}\,{J_r}^{2}+\frac{2}{3}\,{{\it r_c}}^{2
} \left( {\delta E}^{2}{{\it r_c}}^{2}+{J_r}^{2} \right)  \right)  \left( -
\frac{r_c}{2}+M \right) ^{4}{\Omega}^{4}-\\-192\,{\it E_c}\, \left( {M}^{2}+\frac{1}{3}\,M{\it r_c}-\frac{1}{6}\,{{\it r_c}}^{2} \right) {{\it r_c}}^{3} \left( -\frac{r_c}{2}+M \right) ^{3}\delta E\,J_r{\Omega}^{3}-\\-96\, \left( -3\,{M}^{2}{{
\it r_c}}^{4}{\delta E}^{2}+2\,M{{\it r_c}}^{5}{\delta E}^{2}-\frac{1}{3}\,{{\it r_c}
}^{6}{\delta E}^{2}+{J_r}^{2}{M}^{4}-2\,{M}^{3}{\it r_c}\,{J_r}^{2} \right) {
{\it E_c}}^{2}{{\it r_c}}^{2} \left( -\frac{r_c}{2}+M \right) ^{2}{\Omega}^
{2}-\\-576\,{{\it E_c}}^{3}{M}^{3} \left( M-\frac{r_c}{3} \right) {{\it r_c}}^
{5} \left( -\frac{r_c}{2}+M \right) \delta E\,J_r\Omega+240\,{M}^{6}{{\it E_c}
}^{4}{{\it r_c}}^{4}{J_r}^{2}\bigg\rbrace+\mathcal{O}(\varepsilon^5).
\end{multline*}
\end{widetext}

\begin{figure}[h] \centering
	\includegraphics[width=0.85\linewidth]{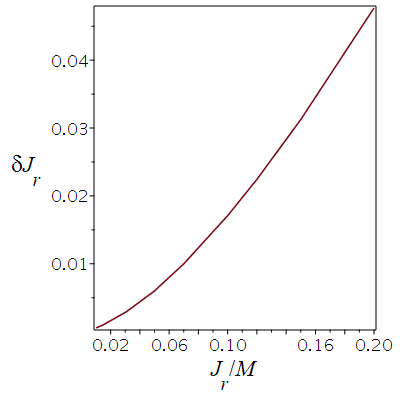}
	\includegraphics[width=0.85\linewidth]{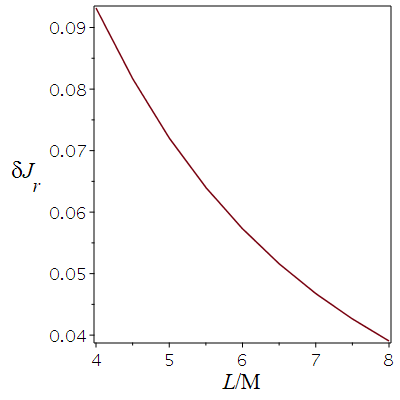}
\caption{ The relative error of the radial action ($\delta J_r$) in the Schwarzschild spacetime for fixed value of  $L= 3.6M$ (top panel) and for $J_r= 0.3M$ (bottom panel).}
\label{fig:deltaJSchw}
\end{figure}

\section{Limits of the approximation} \label{app:tests}

\begin{figure}[h] \centering
	\includegraphics[width=0.95\linewidth]{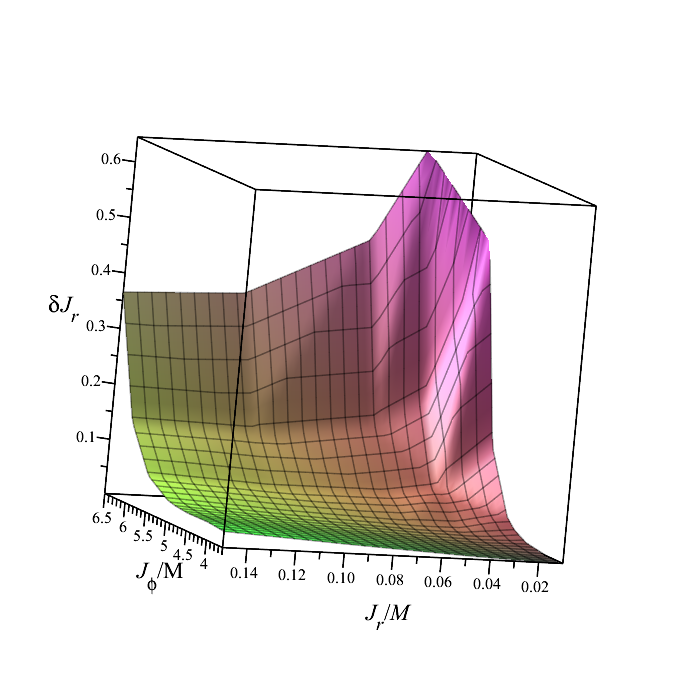}
\caption{ $\delta J_r$ as a function of $J_r$ and $J_\phi$ in the case of equatorial motion for $Q=10^{-6} M^{-2}$.}
\label{fig:deltaJrQ}
\end{figure}

\begin{figure}[h] \centering
	\includegraphics[width=0.85\linewidth]{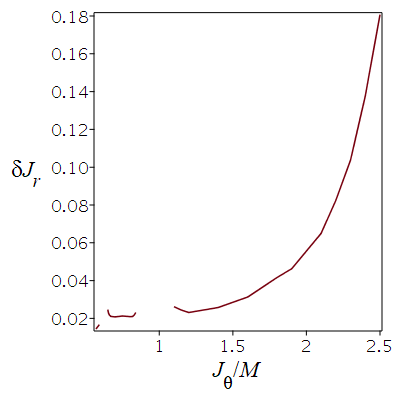}
	\includegraphics[width=0.85\linewidth]{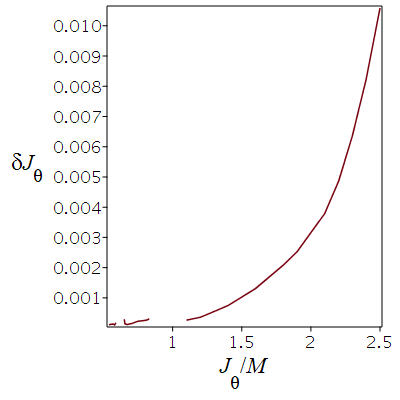}
\caption{Relative error $\delta J_r$ .(top panel) and relative error $\delta J_\theta$ for $J_r =0.1M$, $J_\phi =3M$ and $Q=10^{-6} M^{-2}$ (bottom panel). }
\label{fig:actionreson}
\end{figure}

\begin{figure}[h] \centering
	\includegraphics[width=0.7\linewidth]{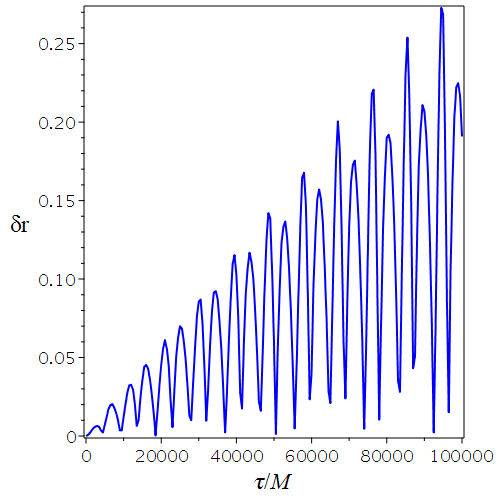}
\caption{Relative error of the radial coordinate as a function of proper time for a non-equatorial orbit}
\label{fig:deltar}
\end{figure}

In the flow generated by $H_N$ the actions $J_r$ and $J_\theta$ are conserved, which is not true in the case of the full Hamiltonian $H$. By inverting the coordinate transformations~\eqref{eq:nc2oc} we can express the actions in terms of our original phase space coordinates. This enables us to evolve actions $J_i$ under $H_{tot}$ and compute the relative error
\begin{align*}
    \delta J_i= \max_{\tau} \frac{\vert J_i^{0}-J_i(\tau)\vert}{J_i^{0}},
\end{align*}
which tells us how  the evolved  actions $J_i(\tau)$ differ from their theoretical counterparts $J_i^{0}$. The maximum is computed for a sufficiently large value of proper time $\tau$, which serves as our evolution parameter here. This relative error is the quantity we will use to test our approximation.

Before switching on the perturbation we should briefly examine the approximation for the Schwarzschild solution itself, for which the situation is fairly simple, since we have only one perturbation parameter $\varepsilon$. The Schwarzschild circular orbits are in our approximation represented exactly and from our expansion of the Schwarzschild Hamiltonian~\eqref{eq:nc2oc}, it is clear that the farther we get from a fixed circular orbit the less accurate our approximation is. This ``phase-space distance'' is measured by the action $J_r$ where $J_r=0$ corresponds to a circular orbit. If our approximation is accurate enough we should approach the separatrix as we increase the value of  $J_r$. When close to the separatrix the approximation should break down which is actually the case. The change of  $\delta J_r$ with  $J_r$ and the total angular momentum $L$ is illustrated in Fig.~\ref{fig:deltaJSchw}.

The value of $L$  selects the circular orbit around which the expansion takes place. The minimum value of $L$, which can be chosen, is $L=2\sqrt{3} M$ representing the ISCO located at $r=6M$. When $L$ is close to its ISCO value, while the value of $J_r$ is large, we can get a bound orbit that can reach the phase-space region corresponding to the infalling  orbits  (for low $r$) leading to a direct contradiction with the exact dynamics. We can, thus, expect that for higher $L$ the approximation is more precise, since the orbit is farther from ISCO, this is verified for example in  Fig.~\ref{fig:deltaJSchw}.

Let us now switch on the perturbation. Even in the case of equatorial motion ($J_\theta=0$) the situation is now slightly more complex than in the Schwarzschild case. As previously the larger the value of $J_r$ the greater the approximation error is. Note that $J_r=0$ still corresponds to circular orbits, however these are not represented exactly \footnote{The exact Schwarzschild circular orbits are shifted by the perturbation.}.  The other relevant action in this setting is $J_\phi$.  The approximation breaks down for large values of $J_\phi$ because we get farther from the black hole and closer to the ring-like source which is equivalent to increasing $Q$.  The dependence $\delta J_r(J_r,J_\phi)$  is illustrated in Fig.~\ref{fig:deltaJrQ}.

In the non-equatorial motion the presence of the resonance of the form  \eqref{eq:reson} has to be taken into account. Thus when plotting the relative errors of actions $\delta J_\theta$ and  $\delta J_r$ as functions of  $J_\theta$ we get an increasing function except for the close neighborhood of the resonance curves where  $\delta J_i \rightarrow \infty$. This corresponds to the two gaps in the graph \ref{fig:actionreson} (resonances $1:2$ and $2:3$).

If we were to plot the functions $r(\tau)$ and $\theta(\tau)$ we would find  that they even leave the domains they are defined on (for example $r<0$). This is caused by the perturbative parts proportional to $Q$ which become larger then the Schwarzschild parts. Alternatively we can plot the relative error of a coordinate as a function of proper time. For example for the radial coordinate we have 
\begin{align}
\delta r(\tau)= \frac{\vert r(\tau)-r^{(n)}(\tau) \vert}{r^{(n)}(\tau)}
\end{align}
where $r^{(n)}(\tau)$ is the numerical solution to the geodesic equation. This error tends to grow over time and even to much larger (but finite) values than the action errors $\delta J_i$. This growth is especially prevalent in the case of non-equatorial  which can be seen in Fig.~\ref{fig:deltar}.

\end{document}